\documentclass{article}
\usepackage{graphicx} % Required for inserting images
\usepackage{subcaption}
\usepackage{amsmath}
\usepackage[english]{babel}
\usepackage{amsmath,amssymb}
\usepackage{graphicx}
\usepackage{booktabs}
\usepackage{hyperref}
\usepackage{color}
\usepackage{multirow}
\usepackage{geometry}
\usepackage{authblk}
\usepackage{hyperref}

\title{Quantum Tunneling and the Emergence of Classical Spacetime in Einstein-Æther Quantum Cosmology}

\author[1]{G. A. Monerat\footnote{E-mail: monerat@uerj.br}}
\author[2]{A. Oliveira Castro J\'{u}nior\footnote{E-mail: alessandroocj@protonmail.com}}
\author[3]{E. V. Corr\^{e}a Silva\footnote{E-mail: profeduvasquezuerj@gmail.com}}
\author[4]{G. Oliveira-Neto\footnote{E-mail: gilneto@fisica.ufjf.br}}
\affil[1,2]{Universidade do Estado do Rio de Janeiro, Instituto Polit\'{e}cnico, Programa de P\'{o}s-Gradua\c{c}\~{a}o em Modelagem Computacional, Nova Friburgo, RJ, Brazil.}
\affil[3]{Departamento de Matem\'{a}tica, F\'{\i}sica e Computa\c{c}\~{a}o, Faculdade de Tecnologia\\
	Universidade do Estado do Rio de Janeiro, CEP 27537-000, Resende - RJ - Brazil.}
\affil[4]{Departamento de F\'{\i}sica, Instituto de Ci\^{e}ncias Exatas\\
	Universidade Federal de Juiz de Fora, Juiz de Fora, Minas Gerais CEP 36036-330, Brazil.}
    
\date{\today}

\begin{document}

\maketitle

\begin{abstract}
We investigate a quantum cosmological model within the framework of Einstein-Æther gravity. The model consists of a positively curved FLRW Universe whose matter content is described by pressureless dust, in the presence of a positive cosmological constant $\Lambda$. We first perform a classical Hamiltonian analysis utilizing Schutz's variational formalism, constructing the phase space and deriving the dynamics of the scale factor. Notably, we identify specific initial conditions that lead to an inflationary expansion, which are extracted directly from the underlying quantum dynamics. Subsequently, we quantize the model following Dirac's formalism, obtaining the Wheeler-DeWitt equation for the wave function of the Universe. We solve this equation numerically and investigate our results using the WKB approximation to compute quantum tunneling probabilities $TP_{WKB}$ alongside an integrated tunneling probability $TP_{int}$, the expectation value of the scale factor coming from the many-worlds interpretation of quantum mechanics, and the corresponding Bohmian trajectories. Finally, we analyze how these tunneling probabilities are influenced by variations in the fundamental parameters of the theory ($\Lambda, \sigma, \beta$) and the energy of the dust fluid $E$. Our results indicate a higher probability for the quantum birth of the Universe in a parameter regime characterized by larger values of $\Lambda, \sigma$, and fluid energy $E$, together with a smaller value of $\beta$. 
\end{abstract}

\section{Introduction}

One of the most profound challenges in contemporary theoretical physics consists in understanding the physical origin of the Universe and establishing a consistent description of its early stages of evolution. Although General Relativity has demonstrated extraordinary success in explaining gravitational phenomena ranging from Solar System scales to galactic dynamics and the large-scale structure of the Universe, its classical formulation is expected to break down under conditions of extreme curvature. In particular, Penrose and Hawking demonstrate that, under fairly general assumptions, cosmological spacetimes described by Einstein's equations inevitably develop geodesic incompleteness. This indicates the existence of an initial singularity where curvature invariants, matter density, and temperature diverge \cite{penrose1965}-\cite{HawkingPenrose1970}. This singular behavior is interpreted as an evidence that the classical description of spacetime becomes incomplete at sufficiently high energies and must eventually be replaced by a quantum theory of gravity \cite{DeWitt1967}-\cite{Bojowald2008}.

\par The quest for a consistent quantum description of the early Universe has motivated the development of various approaches to quantum gravity, including Loop Quantum Gravity \cite{rovelli1990}-\cite{ashtekar2004}, String Theory \cite{green1984}-\cite{maldacena1998}, Causal Dynamical Triangulations \cite{ambjorn2000}-\cite{loll2020}, Asymptotic Safety \cite{reuter1998}-\cite{weinberg1980}, and canonical quantum gravity \cite{DeWitt1967},\cite{dirac1958}-\cite{isham1993} . Among these approaches, quantum cosmology occupies an unique position by providing a simplified framework in which the quantization of the gravitational field can be investigated under the assumptions of spatial homogeneity and isotropy. Instead of quantizing perturbations around a fixed classical background, quantum cosmology seeks to quantize spacetime geometry itself, allowing the Universe to be described by a wave function that obeys the Wheeler-DeWitt equation \cite{DeWitt1967}-\cite{Kiefer2012},\cite{Wheeler1968}-\cite{Kiefer2024}. Although the full superspace possesses an infinite number of gravitational degrees of freedom, physically relevant information can often be extracted from minisuperspace models obtained by imposing spatial homogeneity and isotropy, thus reducing the infinite-dimensional configuration space to a finite set of dynamical variables \cite{Misner1969}-\cite{Kuchar1992}. In recent decades, minisuperspace quantum cosmology has become an extremely productive laboratory for investigating fundamental questions regarding cosmological singularities, the quantum origin of spacetime, the emergence of classical cosmological evolution, and the role of quantum effects during the Planck epoch.

\par A major achievement of Wheeler-DeWitt quantum cosmology is demonstrating that quantum effects regularize spacetime near the initial singularity. Specifically, the scale factor's expectation value and corresponding Bohmian trajectories remain finite, replacing the classical Big Bang with non-singular states, quantum bounces, or tunneling solutions \cite{Bojowald2008}, \cite{HartleHawking1983}-\cite{Ashtekar2006} . These results confirm that early-Universe quantum gravitational mechanisms successfully cure the singular behavior of classical General Relativity.

\par Well-known proposals for the Universe's quantum state include the Hartle-Hawking no-boundary condition \cite{HartleHawking1983}, Vilenkin's tunneling proposal \cite{Vilenkin1982}-\cite{Vilenkin1984}, and Linde's chaotic inflation \cite{Linde1984}. More recently, formulations based on path integrals and Lorentzian approaches have further advanced the understanding of quantum creation and cosmological tunneling processes \cite{Feldbrugge2017}-\cite{DiazDorronsoro2018}."

\par Parallelly, advances in numerical quantum cosmology now allow tracking wave packet evolution and computing tunneling probabilities directly from the Wheeler-DeWitt equation, bypassing semiclassical approximations. Probability-conserving finite-difference algorithms, such as Crank-Nicolson, have been successfully applied to various scenarios \cite{Monerat2007}-\cite{OliveiraNeto2011} , recently proving their versatility in Lorentz-violating and Einstein-Æther models \cite{Alessandro2026}.

\par Despite these important advances, one of the central conceptual issues in quantum cosmology remains largely unresolved. Most existing investigations focus primarily on demonstrating that quantum effects eliminate the initial singularity or allow the Universe to tunnel through an effective potential barrier. However, relatively little attention has been devoted to understanding how the classical Universe effectively emerges from the underlying quantum dynamics. In many minisuperspace models, the quantum and classical analyses are performed independently: after solving the Wheeler-DeWitt equation, one simply assumes a set of classical initial conditions and investigates the subsequent cosmological evolution \cite{AlvarengaLemos1998,Monerat2007,Vakili2010}. Although this procedure has been extremely useful for studying classical cosmological solutions, it leaves unanswered one of the most fundamental questions in quantum cosmology: how are the initial conditions that govern the classical Universe determined by the preceding quantum evolution?

\par The quantum-classical transition thus constitutes one of the most fundamental open problems in quantum cosmology. If the Universe indeed originated in a genuinely quantum regime, one expects the classical spacetime to be the natural evolution of it. This issue has been extensively investigated \cite{Kiefer2012}, \cite{Zeh1970}-\cite{Halliwell1999}, in particular, decoherence induced by inaccessible spatial degrees of freedom provides a compelling mechanism by which classical correlations can emerge from quantum superpositions without requiring any modification of the underlying quantum dynamics. However, despite substantial conceptual progress, a fully self-consistent implementation of the quantum-classical transition within explicit minisuperspace models remains relatively rare.

\par Motivated by this gap, we propose a unified framework rather than treating the quantum and classical regimes as disconnected. We use expectation values derived from the numerical Wheeler-DeWitt evolution to set the initial conditions for the post-tunneling classical cosmology. Consequently, the onset of the classical Universe emerges directly from the quantum dynamics rather than being phenomenologically imposed, providing a continuous, self-consistent description from the primordial state to the classical expansion.

\par To extend General Relativity beyond its classical domain, theories with spontaneous breaking of local Lorentz invariance have gained considerable traction. While Lorentz symmetry is well tested at current energies, it may simply be a low-energy effective symmetry of fundamental quantum gravity. Consequently, Lorentz-violating scenarios naturally emerge in frameworks like Hořava-Lifshitz gravity, Einstein-Æther theory, the Standard-Model Extension, and various effective field theories \cite{Jacobson2001}-\cite{Liberati2013}.

\par Einstein-Æther gravity stands out among these alternatives by dynamically breaking local Lorentz invariance through a unit timelike vector field which is known as "the aether" while preserving general covariance and diffeomorphism invariance \cite{Jacobson2001}-\cite{Jacobson2004}. Because its equations remain second-order in metric derivatives, it successfully avoids the instabilities typical of higher-order modified gravity, establishing it as a leading Lorentz-violating framework. More recently, the direct detection of gravitational waves by the LIGO-Virgo-KAGRA Collaboration has placed strong observational constraints on Lorentz-violating theories, significantly reducing the allowed parameter space of Einstein-Æther gravity, although phenomenologically viable regions compatible with current cosmological observations still remain \cite{oost2018}-\cite{abbott2017}.

\par Previous investigations have demonstrated that the aether field can significantly affect inflationary dynamics, bouncing cosmologies, dark energy evolution, and the growth of primordial perturbations, suggesting that effects associated with Lorentz violation may play an important role during the earliest stages of cosmic evolution \cite{Lim2005,Carroll2004,Donnelly2010,Audren2013,oost2018,Lim2005}.

\par The quantum cosmology of Einstein-Æther models remains largely unexplored. Specifically, the impact of Lorentz-violating couplings on quantum nucleation, tunneling probabilities, and the emergence of classical initial conditions lacks a systematic Wheeler-DeWitt treatment. Bridging this gap is the primary motivation of the present work.

\par Our recent investigations demonstrate that Lorentz-violating corrections in Einstein-Æther gravity significantly alter the classical and quantum dynamics of early-Universe models involving radiation, dark energy, or cosmic strings \cite{Alessandro2026},\cite{castrojunior2026}. Because these effects dominate at high energies, extending this framework is highly motivated. To this end, we consider an FLRW Universe filled with pressureless dust \cite{nelson2026} and a positive cosmological constant. Although standard models in cosmology assumes early radiation or scalar-field domination, the use of dust in canonical quantum cosmology possesses both formal and physical motivations, making it particularly well-suited for investigating the quantum-classical transition.

\par From a formal point of view, pressureless dust described through Schutz's variational formalism plays a fundamental role in canonical quantum cosmology, as it naturally introduces an intrinsic time variable into the Hamiltonian formulation \cite{schutz1970}-\cite{lapchinskii1977}. %The canonical momentum associated with the fluid appears linearly in the Hamiltonian constraint, allowing the Wheeler-DeWitt equation to assume the mathematical structure of a Schrödinger-like evolution equation. This remarkable property provides a well-defined notion of quantum evolution and, consequently, has made Schutz's formalism one of the standard approaches for dealing with the long-standing problem of time in canonical quantum gravity.

In recent decades, this formalism has been successfully applied to isotropic and anisotropic cosmologies, Chaplygin gas models, modified theories of gravity, and numerous cosmological scenarios involving quantum bounces (bouncing cosmologies) \cite{AlvarengaLemos1998,Monerat2007,Pedram2008,Vakili2010,OliveiraNeto2011,monerat2025}.

\par Beyond mathematical convenience, pressureless matter ($w=0$) is physically relevant in the early Universe. Several mechanisms, including coherently oscillating massive scalar fields, moduli fields, curvatons \cite{lyth2002},\cite{lyth2003},  reheating dynamics, and primordial black holes \cite{turner1983}-\cite{carr2020}, generate effective dust-dominated epochs. Although the model considered in this work does not aim to describe reproduce the complete thermal historyof the Universe, pressureless dust simultaneously provides a mathematically consistent internal clock and effectively captures these physically motivated early-Universe phases.

The primary objective of the present work is not simply to investigate yet another minisuperspace solution to the Wheeler-DeWitt equation, but rather to clarify how Lorentz-violating gravitational dynamics can influence both the quantum origin of the Universe and its subsequent emergence as a classical cosmological spacetime.

\par In contrast to most previous studies, in which the quantum and classical descriptions are treated independently, we seek to establish a unified framework in which the classical evolution emerges continuously from the preceding quantum dynamics. %Our central hypothesis is that the information required to initialize the classical Universe can be extracted directly from the quantum evolution after the tunneling process, thus providing a self-consistent realization of the quantum-classical transition within Einstein-Æther cosmology.

To achieve this goal, we first investigate the full classical dynamics of an FLRW Universe in Einstein-Æther gravity filled with dust and a positive cosmological constant. Starting from the minisuperspace Hamiltonian obtained through Schutz's variational formalism, we analyze the corresponding effective potential and classify all admissible classical solutions according to their total energy.

This analysis reveals that the cosmological dynamics naturally separates into four distinct classes of solutions, including eternally expanding Universes, recollapsing solutions, and bouncing cosmologies, whose existence is determined by the structure of the effective potential generated by the Lorentz-violating gravitational couplings.

\par Next, we formulate the corresponding Wheeler-DeWitt equation and solve it numerically using the implicit Crank-Nicolson finite-difference method. Unlike semiclassical treatments based on the WKB approximation, the numerical procedure adopted here tracks the complete evolution of the quantum wave packet during the tunneling process, allowing the tunneling probability to be calculated directly from the exact quantum dynamics, without introducing additional approximations.

This approach makes it possible to determine not only the probability that the Universe emerges beyond the potential barrier, but also the detailed evolution of the quantum state as it propagates through classically forbidden regions.

A primary innovation of this work is explicitly bridging the quantum and classical regimes. We extract the classical initial conditions directly from the numerical quantum evolution. Specifically, expectation values calculated immediately after the wave packet exits the classically forbidden region serve as initial data for the classical Hamiltonian equations. Thus, classical cosmology emerges as a natural continuation of the quantum dynamics rather than an independent stage.

In addition to providing a conceptually consistent realization of the quantum-classical transition, this procedure also offers a new perspective on the physical interpretation of inflationary cosmology. In conventional inflationary scenarios, the onset of inflation is generally assumed following the specification of appropriate classical initial conditions \cite{guth1981}-\cite{baumann2011}. However, in the framework developed in this work, these initial conditions are not externally imposed, but emerge from the quantum dynamics itself.

\par Consequently, the inflationary phase can be interpreted as a direct consequence of the quantum birth of the Universe, rather than an independent hypothesis introduced after quantization. Establishing a connection between quantum cosmology and the subsequent classical accelerated expansion constitutes one of the primary conceptual motivations of this work.

\par A second objective of this investigation is to systematically analyze the influence of the Einstein-Æther coupling parameters on the quantum nucleation of the Universe. By calculating the tunneling probability for a wide range of energies and Lorentz-violating coupling constants, we examine how the modified gravitational dynamics affects the quantum creation probability, and we compare the exact numerical results with the corresponding semiclassical WKB approximation.

This comparison allows us to identify the physical origin of the discrepancies observed between both approaches and to determine the parameter regimes in which semiclassical methods provide reliable approximations to the exact quantum dynamics.

\par We analyze the quantum solutions using two complementary frameworks: the Many-Worlds interpretation to compute expectation values, and the de Broglie-Bohm formulation to construct quantum trajectories. Despite their conceptual differences, both approaches yield the same conclusion: the scale factor remains strictly finite throughout the evolution. This agreement robustly confirms the replacement of the classical singularity with regular quantum dynamics.

\par In summary, this study advances quantum cosmology in three complementary directions: (i) incorporating Lorentz-violating dynamics into an FLRW Einstein-Æther Universe; (ii) establishing an explicit quantum-classical transition where classical initial conditions emerge strictly from numerical Wheeler-DeWitt solutions; and (iii) systematically detailing how aether couplings dictate both quantum tunneling and subsequent classical evolution. Ultimately, this yields a unified, self-consistent framework bridging the Universe's quantum birth and classical expansion.

\par This article is structured as follows: in Section \ref{sec:classical_model}, we build the classical model, obtain its classical scale factor solutions, and analyze the inflationary behavior of the model. Section \ref{sec:quantum_model} is dedicated to quantizing the model, deriving the Wheeler-DeWitt equation, and solving it to yield the complete wave function $\Psi(x,\tau)$. Then, we use the Many-Worlds and de Broglie-Bohm interpretations to calculate the scale factor expectation value and the Bohmian trajectory, respectively, and compare them. In Section \ref{sec:classical_quantum_transition}, we show that it is possible to obtain a classical inflationary model starting from initial conditions derived from the quantum version of the model. Finally, in Section \ref{sec:tunneling_probs}, we study how the Einstein-Æther parameters, as well as the fluid energy, may affect the tunneling probabilities.

%-------------------------------------------------------------------------------------------------------

%-------------------------------------------------------------------------------------------------------

\section{The structuring of the Classical Model}
\label{sec:classical_model}

We take as our starting point the Einstein--Hilbert action combined with the Gibbons--Hawking--York (GHY) boundary term, given by
\begin{equation}
    S_{EH} = \frac{1}{16\pi G} \left[ \int_M d^4x \sqrt{-g} \, R + 2 \int_{\partial M} d^3x \sqrt{h} \, K \right],
    \label{eq:EH}
\end{equation}

\noindent where \(G\) is Newton's gravitational constant and \(R\) denotes the Ricci scalar of the spacetime metric \(g_{\mu\nu}\) on the four-dimensional manifold \(M\), with determinant \(g \equiv \det(g_{\mu\nu})\). The boundary \(\partial M\) is a three-dimensional spacelike hypersurface equipped with the induced metric \(h_{ab}\) (whose determinant is \(h\)), and \(K_{ab}\) is the extrinsic curvature of that hypersurface. The quantity \(K \equiv h^{ab}K_{ab}\) is the trace of the extrinsic curvature, which enters the boundary action to ensure that the variational principle for the gravitational field is well-posed with Dirichlet boundary conditions on the induced metric.

Now we proceed with the construction of the complete action by introducing the kinetic sector for the aether field $u^a$. This field is taken to be a unit timelike vector, satisfying the constraint $u^a u_a = -1$ adopting the $(-+++)$ signature convention. The dynamics of this sector are governed by terms that depend on the field itself and on its first covariant derivatives. In the present model, these contributions are encoded in the generalized kinetic action \cite{Jacobson2008AEreport}
\begin{equation}
    S_{AE} = \frac{1}{16\pi G} \int_M d^4x \sqrt{-g} \left( -B_{mn}^{ab} \nabla_a u^m \nabla_b u^n + \lambda (u^a u_a + 1) \right),
\label{eq:SAE}
\end{equation}

\noindent where the generalized coupling tensor is defined as

\begin{equation}
    B_{mn}^{ab} = c_1 g^{ab} g_{mn} + c_2 \delta^a_m \delta^b_n + c_3 \delta^a_n \delta^b_m + c_4 u^a u^b g_{mn}.
\end{equation}

\noindent The generalized coupling tensor $B_{mn}^{ab}$ introduced above is constructed from the four dimensionless constants $c_i$ (with $i = 1,\dots,4$), which parametrize the strength of the interaction between the aether field $u^a$ and the spacetime metric $g_{ab}$. The scalar field $\lambda$ acts as a Lagrange multiplier whose role is to enforce the unit-timelike constraint $u^a u_a = -1$ (equivalently, $u^a u_a + 1 = 0$ in the action), thereby fixing the norm of the aether vector throughout the spacetime manifold. In the limiting case where all $c_i$ vanish identically, the tensor $B_{mn}^{ab}$ reduces to zero, the aether kinetic sector decouples entirely from the gravitational dynamics, and the total action consistently reduces to that of General Relativity with matter fields.

We consider a homogeneous and isotropic cosmological background described by the Friedmann--Lema\^{\i}tre--Robertson--Walker (FLRW) metric. This universe is assumed to be filled with a barotropic perfect fluid, whose equation of state is given by $p = \alpha \rho$, where $p$ denotes the isotropic pressure and $\rho$ is the corresponding energy density. The constant parameter $\alpha$, which characterizes the fluid species, is restricted to the range $-1 \leq \alpha \leq 1$ for standard thermodynamically stable fluids. Specifically, $\alpha = 0$ corresponds to non-relativistic pressureless matter (dust), $\alpha = 1/3$ describes an ultra-relativistic fluid (radiation), and $\alpha = -1$ yields a cosmological constant. For completeness, we note that the case $\alpha < -1$ defines a phantom fluid, although such values lie outside the standard interval considered in the present work.

To describe the dynamics of the perfect fluid, we adopt Schutz's variational formalism \cite{Schutz1,Schutz2,Furtado}, in which the fluid four-velocity $U_\nu$ is expressed in terms of the thermodynamic potentials $\epsilon$, $\theta$, $S$, $\zeta$, and $W$. Given the homogeneity and isotropy of the FLRW background, the potentials $\zeta$ and $W$ must vanish identically, as they would otherwise introduce preferred spatial directions or inhomogeneities. We therefore set $\zeta = W = 0$, which reduces the fluid velocity to the simpler form
\begin{equation}
U_\nu = \frac{1}{\mu}(\epsilon_{,\nu} + \theta s_{,\nu}) ,
\label{eq:U_nu_reduced}
\end{equation}
consistent with the definition introduced previously.

\noindent In the reduced expression for the fluid four-velocity, $\mu$ denotes the specific enthalpy and $s$ is the specific entropy. The remaining potentials $\epsilon$ and $\theta$, by contrast, are auxiliary variables introduced within Schutz's formalism and do not possess a direct physical interpretation; they typically encode the velocity potential decomposition and are eliminated via the variational equations. With these ingredients, Schutz's variational procedure yields the final component of our total action, namely the matter sector, which takes the form
\begin{equation}
    S_M = \int_M d^4x \, \sqrt{-g} \, p .
    \label{eq:SM}
\end{equation}
This completes the construction of the full action, given by the sum $S = S_{EH} + S_{AE} + S_M$ as presented in Eqs.~\eqref{eq:EH},~\eqref{eq:SAE}, and~\eqref{eq:SM} above.

\noindent The complete action for our model now reads

\begin{equation}
\begin{aligned}
    S &= S_{EH} + S_{AE} + S_M \\
    &= \frac{1}{16\pi G} \left[ \int_M d^4x \sqrt{-g} R + 2 \int_{\partial M} d^3x \sqrt{h} \, K \right] + \\
    &\frac{1}{16\pi G} \int_M d^4x \sqrt{-g} \left( -B_{mn}^{ab} \nabla_a u^m \nabla_b u^n + \lambda (u^a u_a + 1) \right) + \\
    &\int_M d^4x \, \sqrt{-g} \, p
\label{full_action}
\end{aligned}
\end{equation}

We adopt the spatially homogeneous and isotropic Friedmann--Lema\^{\i}tre--Robertson--Walker (FLRW) line element, written in spherical coordinates as
\begin{equation}
ds^2 = -{N(t)}^2dt^2 + a(t)^2 \left[ \frac{dr^2}{1 - k r^2} + r^2 \left( d\theta^2 + \sin^2\theta \, d\phi^2 \right) \right],
\label{eq:FLRW}
\end{equation}
where $t$ is the cosmic time, $N(t)$ is the lapse function and $a(t)$ is the cosmic scale factor, where $k = -1, 0, +1$ corresponds to open, flat, and closed spatial geometries, respectively. 

We now reduce the total action in Eq.~\eqref{full_action}. The last term in in Eq.~\eqref{full_action} represents the fluids to be introduced. This term will be the origin of a vacuum fluid whose equation of state is $p_\Lambda=-\rho_\Lambda$ which plays the role of a dark energy. Also, a second fluid which will play the role of a pressureless dust fluid whose equation of state is $p=0$. Following the thermodynamic prescription of Lapchinskii and Rubakov \cite{lapchinskii1977}, which accounts for the fluid constraints, and applying the Einstein-Æther reduction scheme of Campista et al.\cite{campista2020}, while discarding all surface contributions, the action reduces to 

\begin{equation}
    \mathcal{S} = \int dt \left[ -\frac{6a\dot{a}^2}{\sigma} - \frac{3\beta a}{\sigma} \dot{a}^2 + \frac{a^3}{\sigma} \frac{\alpha(\dot{\epsilon} + \theta\dot{s})^{1/\alpha+1}}{(\alpha+1)^{1/\alpha+1}} e^{-\frac{s}{\alpha}} \right],
\label{eq:reduced_action}
\end{equation}

\noindent where \(a(t)\) is the scale factor and \(\sigma\) and \(\beta\) are combinations of the aether coupling constants \(c_i\):

\begin{equation}
    \sigma = 1 - \frac{c_1 + c_4}{2}, \hspace{0.5cm} \beta = c_1 + 3c_2 + c_3.
\end{equation}

\noindent For $\sigma = 1$ and $\beta = 0$, GR is recovered. A Legendre transformation of the reduced minisuperspace Lagrangian yields the Hamiltonian

\begin{equation}
    H = N \mathcal{H} = \frac{1}{2m} {p_a}^2 + \frac{m}{2}{\omega_k}^2 a^2 - \Lambda a^{1-3\Xi} - a^{1-3\alpha} p_T,
\end{equation}

\noindent where \(H\) denotes the total Hamiltonian and \(\mathcal{H}\) is the super-Hamiltonian, which is subject to the standard constraint \(\mathcal{H}=0\). For the present work, $\Xi = -1$ representing a dark energy term whereas the parameter $\alpha$ will be set as $\alpha = 0$. The parameters appearing in the Hamiltonian are related to the fundamental constants of the aether sector through
\begin{equation}
m = \frac{6(\beta + 2)}{\sigma}, \qquad 
\omega_k = \sqrt{\frac{2k}{\beta + 2}},
\label{eq:theory_parameters}
\end{equation}

\noindent In particular, $\sigma$ encodes the overall strength of the aether modifications, while $\beta$ encapsulates the anisotropic contributions. The parameter $m$ plays the role of an effective mass in the minisuperspace Hamiltonian, and $\omega_k$ corresponds to the effective frequency associated with the spatial curvature of the FLRW geometry. The lapse function \(N\) encodes the residual gauge freedom associated with time-reparametrization invariance. In what follows, we fix this gauge by adopting the conformal gauge, \(N = a\). We added a term containing $\Lambda$, taking into account the cosmological constant influence on our model. With this choice, the FLRW line element becomes
\begin{equation}
ds^2 = a(\eta)^2 \left[ -d\eta^2 + \frac{dr^2}{1 - k r^2} + r^2 \left( d\theta^2 + \sin^2\theta \, d\phi^2 \right) \right],
\label{eq:FLRW_conformal}
\end{equation}
\noindent In this gauge, the coordinate \(\eta\) is explicitly identified with conformal time, and the overall factor \(a(\eta)^2\) serves as the conformal factor of the metric. This choice significantly simplifies the structure of the Hamiltonian and is standard in minisuperspace analyses of quantum cosmology. We may define

\begin{equation}
    V_{eff} = \frac{m}{2}{\omega_k}^2 a^2 - \Lambda a^4
\end{equation}

such that one can write the Hamiltonian as 

\begin{equation}
    H = \frac{1}{2m} {p_a}^2 + V_{eff}(a) - a^{1-3\alpha} p_T.
\end{equation}

\begin{figure}[!htpb]
    \centering
    \includegraphics[scale=0.28]{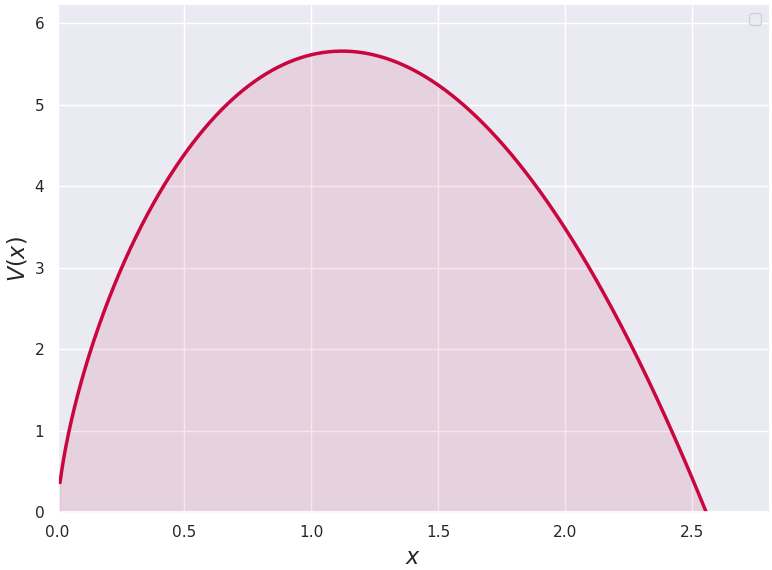}
    \caption{Potential $V(x)$. The chosen values for the parameters are $k=1, \Lambda = 1.0, \beta = 0.27$ and $\sigma =0.9999$. The values for $\beta$ and $\sigma$ were obtained from observational data described in Physical Review D, 97, 124023, (2018).}
    \label{potential}
\end{figure}

To address the well-known factor-ordering ambiguities that arise upon canonical quantization of the minisuperspace Hamiltonian, we introduce a change of variables from the scale factor $a(t)$ to a new coordinate $x(t)$, defined through the relations \cite{silva2009}
\begin{align}
    a &= \left( \frac{3(1-\alpha)x}{2} \right)^{\frac{2}{3-3\alpha}}|_{\alpha=0} = \left( \frac{3x}{2} \right)^{2/3}, \\
    p_a &= p_x \, a^{\frac{1-3\alpha}{2}}|_{\alpha=0} = p_x a^{1/2}, \\
    F &= \frac{2}{3(1-\alpha)} p_x \, a^{\frac{3-3\alpha}{2}}|_{\alpha=0} = \frac{2}{3}p_x a^{3/2} ,
\end{align}
\noindent where $F$ is the generating function. For the specific case of a dust-dominated universe, which corresponds to $\alpha = 0$, the above transformation reduces the Hamiltonian to the simpler form
\begin{equation}
    H = \frac{p_x^2}{2m} + V_{\text{eff}}(x) - p_T,
    \label{eq:H_dust}
\end{equation}
where the effective potential now reads
\begin{equation}
    V_{\text{eff}}(x) = \frac{m}{2} \omega_k^2 \left( \frac{3}{2}x \right)^{2/3} - \Lambda \left( \frac{3}{2}x \right)^2,
    \label{eq:V_eff_dust}
\end{equation}

\noindent and it is shown in Fig.\ref{potential}.

From the Hamiltonian in Eq.~(\ref{eq:H_dust}), Hamilton's equations yield the following system of first-order differential equations:

\begin{equation}
\left\{
\begin{array}{lllllll}
\dot{x}=&\frac{p_x}{m} ;\\
&\\
\dot{p_{x}}=&- \frac{m}{3} \omega_k^2 \left( \frac{3}{2} \right)^{2/3} x^{-1/3} + \frac{9}{2} \Lambda x;\\
&\\
\dot{T}=&\frac{\partial {H}}{\partial p_{T}}=-1 \, ; \\
&\\
\dot{p_{T}}=&-\frac{\partial {H}}{\partial T}=0.
\end{array}
\right.
\label{hamilton_eqs}
\end{equation}

\noindent where the overdot denotes differentiation with respect to conformal time $\eta$. The equation \(\dot{T} = -1\) integrates to \(T = -\eta + \text{const.}\), reflecting the standard behaviour of the Schutz time variable.  Similarly, \(\dot{p}_T = 0\) implies that \(p_T\) is a constant of motion; physically, it corresponds to the conserved energy (or mass) of the dust fluid within the Schutz formulation. 

Imposing the Hamiltonian constraint \(H = 0\) yields the algebraic relation
\begin{equation}
    \frac{p_x^2}{2m} = -V_{\text{eff}}(x) + p_T ,
    \label{eq:constraint}
\end{equation}
where \(V_{\text{eff}}(x)\) is given by Eq.~(\ref{eq:V_eff_dust}). The critical value of the conserved quantity \(p_T\), denoted \(p_{T,\text{crit}}\), corresponds to the extremum of the effective potential and is obtained by setting \(p_x = 0\). This leads to the condition
\begin{equation}
    \frac{m}{2} \omega_k^2 \left( \frac{3}{2}x \right)^{2/3} - \Lambda \left( \frac{3}{2}x \right)^2 - p_T = 0 .
    \label{eq:critical_condition}
\end{equation}
Solving for the extremum of the potential, we obtain the critical value
\begin{equation}
    p_{T\text{crit}} = \frac{m^{3/2} \omega_k^3}{3\sqrt{6\Lambda}} .
    \label{eq:pT_crit}
\end{equation}

To derive the classical equation of motion for the scale factor, we start from the constraint equation, which gives the canonical momentum as
\begin{equation}
    p_x = \sqrt{ 2m \left[ p_T - \frac{m}{2} \omega_k^2 \left( \frac{3}{2}x \right)^{2/3} + \Lambda \left( \frac{3}{2}x \right)^2 \right] } .
    \label{eq:px_sqrt}
\end{equation}
Using the Hamilton equation \(\dot{x} = p_x / m\), we have \(p_x = m \dot{x}\). Substituting this into Eq.~(\ref{eq:px_sqrt}) and differentiating with respect to \(\eta\) yields the second-order equation of motion
\begin{equation}
    \ddot{x} = \frac{9\Lambda}{2m}x - \frac{\omega_k^2}{2}\left( \frac{3x}{2} \right)^{-1/3} .
    \label{eq:second_order_x}
\end{equation}

\begin{figure}[!htpb]
    \centering
    \includegraphics[scale=0.3]{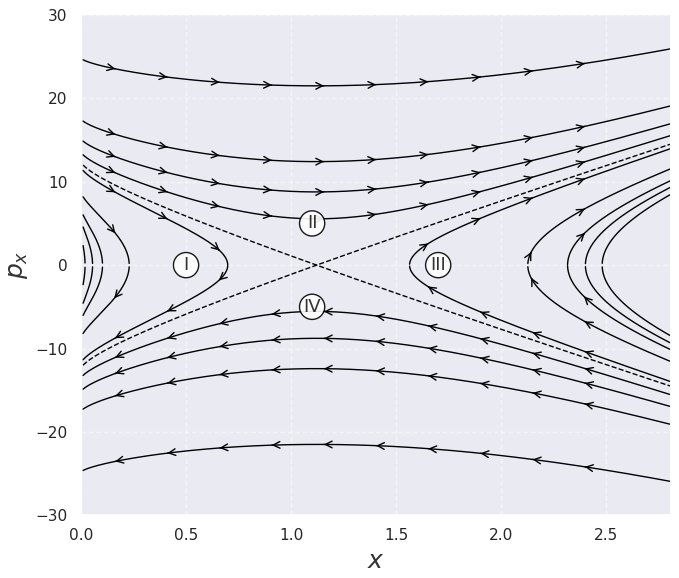}
    \caption{Phase Space $(x,p_x)$. The separatrices correspond to the critical value $p_{T\text{crit}}$ from Eq.~(\ref{eq:pT_crit})}
    \label{phase_space}
\end{figure}

\begin{figure}[!htpb]
    \centering
    \includegraphics[scale=0.35]{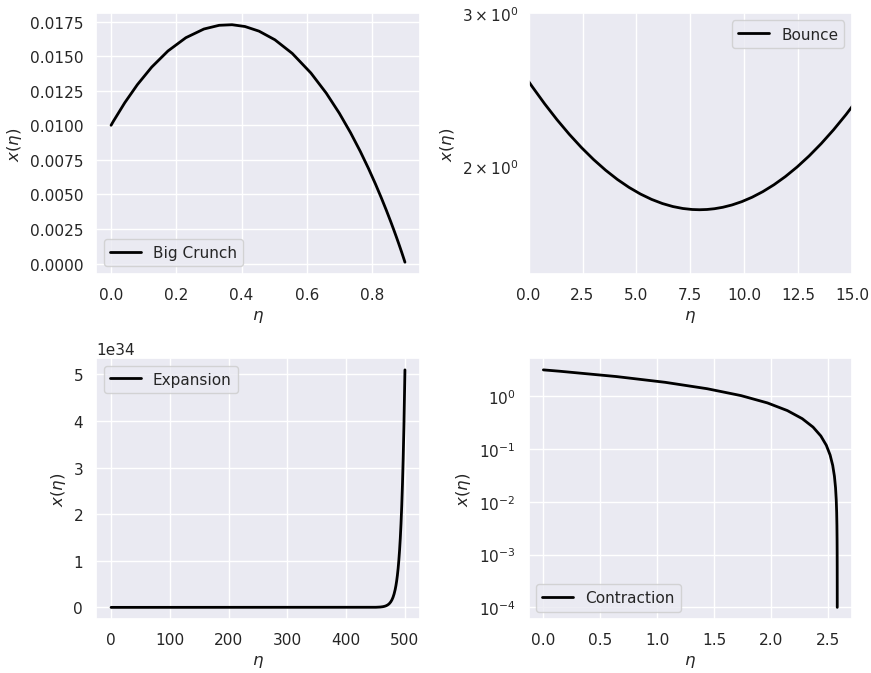}
    \caption{Classical Scale factor solutions}
    \label{scale_factor}
\end{figure}

Depending on the choice of initial conditions, the classical dynamics described by Eq.~(\ref{eq:second_order_x}) yield four qualitatively distinct families of solutions, corresponding to four separate regions in the $(x, p_x)$ phase space. These regions are illustrated in Fig.~\ref{phase_space} and may be characterised as follows: (I) big crunch, (II) expansion, (III) bounce, and (IV) contraction. The boundaries separating adjacent regions, indicated by dashed lines in Fig.~\ref{phase_space}, are referred to as separatrices. Representative solutions from each of the four classes are displayed in Fig.~\ref{scale_factor}, which exemplifies the typical behaviour of the scale factor within each regime.

To illustrate the four types of dynamical behaviour, we solve Eq.~(\ref{eq:second_order_x}) numerically for the following sets of initial conditions:
\begin{itemize}
    \item \textbf{Big crunch (region I):} $x(0) = 0.01$, $\dot{x}(0) = 0.0415355$;
    \item \textbf{Expansion (region II):} $x(0) = 0.1$, $\dot{x}(0) = 2.49213$;
    \item \textbf{Bounce (region III):} $x(0) = 2.5$, $\dot{x}(0) = -0.20767$;
    \item \textbf{Contraction (region IV):} $x(0) = 3.12$, $\dot{x}(0) = -1.2460$.
\end{itemize}
Each set of initial conditions yields a solution that lies entirely within the corresponding phase-space region, as displayed in Fig.~\ref{scale_factor}.

\subsection{The inflationary behavior}
The concept of cosmic inflation was introduced in the early 1980s to resolve several discrepancies between theoretical predictions and observational data \cite{inflation1,inflation2,inflation3}, as well as to address certain foundational issues in standard Big Bang cosmology. The central idea is that the very early Universe underwent a period of exponential expansion, after which it transitioned to the slower, decelerated expansion observed today. This framework finds strong observational support in Hubble's law, formulated in 1929, which establishes that distant galaxies recede from one another with a recession velocity proportional to their distance.

With this empirical motivation, we now examine whether our model admits configurations capable of reproducing the salient features of inflation. Specifically, the standard inflationary paradigm holds that the expansion occurred from approximately \(10^{-36}\) s to \(10^{-32}\) s after the Big Bang, during which the scale factor grew by a factor of order \(10^{26}\), with a number of e-folds
\cite{WeinbergCosmology}
\begin{equation}
	D = ln\left( \frac{a_f}{a_i} \right) \approx 60.
\end{equation}

\noindent In our phase-space analysis, regions II and III in Fig.~\ref{phase_space} correspond to expanding and bouncing solutions, respectively. We therefore focus on initial conditions within these regions that may yield an exponential growth of the scale factor consistent with the following observational benchmarks:
\begin{itemize}
    \item a ratio of the scale factor at the end of inflation to that at the onset of approximately \(10^4\);
    \item a total expansion factor of the Universe of order \(10^{26}\);
    \item a sufficient number of e-folds of inflation, typically \(N \approx 60\), to resolve the horizon and flatness problems.
\end{itemize}
By systematically exploring the parameter space of our model, we seek to identify data that satisfy these criteria, thereby providing a viable inflationary scenario within the present framework. Table~\ref{tab:inflationary_solutions} shows calculations of e-folds values for the initial conditions mentioned before regarding expansion and bouncing solutions. Corresponding time intervals computed in regions II and III yields expansion ratios of $\approx 10^{26}$ . We have expressed our time intervals in terms of the conformal time $\eta$. We have defined $N_x \equiv ln(x_{final}/x_{ini})$ and we also did the transformation back to check the e-folds in the original scale factor $a(\eta)$ through a quantity we defined $N_a \equiv ln(a_{final}/a_{ini})$.

{ \small
\begin{table}[htbp]
\centering
\hspace*{-0.2cm}
\begin{tabular}{|l|c|c|r|r|r|r|}
\hline
\textbf{Type} & \textbf{\(\eta_{\text{ini}}\)} & \textbf{\(\eta_{\text{final}}\)} & \textbf{Ratio ($x_{final}/x_{ini}$)} & \textbf{e-folds (\(N_x\))} & \textbf{Ratio ($a_{final}/a_{ini}$)} & \textbf{e-folds (\(N_a\))} \\ \hline

Expansion & 0.01 & 1000.0 & \(2.9995 \times 10^{40}\) & 93.2018 & \(9.6537 \times 10^{26}\) & 62.1346 \\ \hline
Bounce    & 0.05 & 5000.0 & \(1.8358 \times 10^{40}\) & 92.7109 & \(6.9591 \times 10^{26}\) & 61.8073 \\ \hline
\end{tabular}
\caption{Numerical results for the expanding and bouncing solutions. The quantities \(\eta_{\text{ini}}\) and \(\eta_{\text{final}}\) denote the initial and final conformal times, respectively. The ratios and corresponding e-folding numbers are given for both the transformed variable \(x\) and the scale factor \(a\).}
\label{tab:inflationary_solutions}
\end{table}
}

\subsection{The Hubble parameter}

The Hubble parameter \cite{hubble1,hubble2,deceleration}, expressed as a function of cosmic time $t$, is defined as
\begin{equation}
    H_b(t) \equiv \frac{\dot{x}(t)}{x(t)},
\end{equation}
while the deceleration parameter \cite{deceleration} is given by
\begin{equation}
    q(t) \equiv -\frac{\ddot{x}(t) x(t)}{\dot{x}(t)^2}.
\end{equation}
These quantities satisfy the relation
\begin{equation}
    \dot{H}_b = -H_b^2(1+q) \iff q = -\left(\frac{\dot{H}_b}{H_b^2} + 1\right),
\label{eq:relation_H_q}
\end{equation}
where overdots denote differentiation with respect to $t$. In the standard $\Lambda$CDM paradigm, as the cosmological constant becomes the dominant energy component, the deceleration parameter tends to $q \to -1$. Consequently, Eq.~(\ref{eq:relation_H_q}) implies $\dot{H}_b \to 0$, so that the Hubble parameter approaches a constant value and the scale factor grows exponentially with time.

%In conformal time $\eta$, where a prime denotes differentiation with respect to $\eta$, the above definitions translate into
%\begin{equation}
%    H_b(\eta) = \frac{a'(\eta)}{a(\eta)^2},
%\end{equation}
%and
%\begin{equation}
%    q(\eta) = 1 - \frac{a(\eta) \, a''(\eta)}{\left[a'(\eta)\right]^2}.
%\end{equation}
%As the qualitative behaviour of these parameters is invariant under the time reparametrization, we shall employ the conformal-time expressions to analyse our model.

\begin{figure}
    \centering
    \includegraphics[scale=0.4]{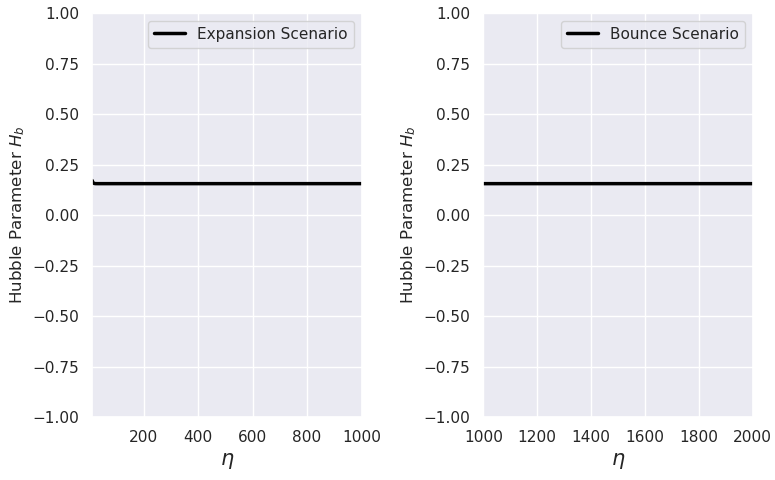}
    \caption{Hubble parameter for the scenarios depicting the expansion and bounce solutions.}
    \label{hubble_parameter}
\end{figure}

\begin{figure}
    \centering
    \includegraphics[scale=0.4]{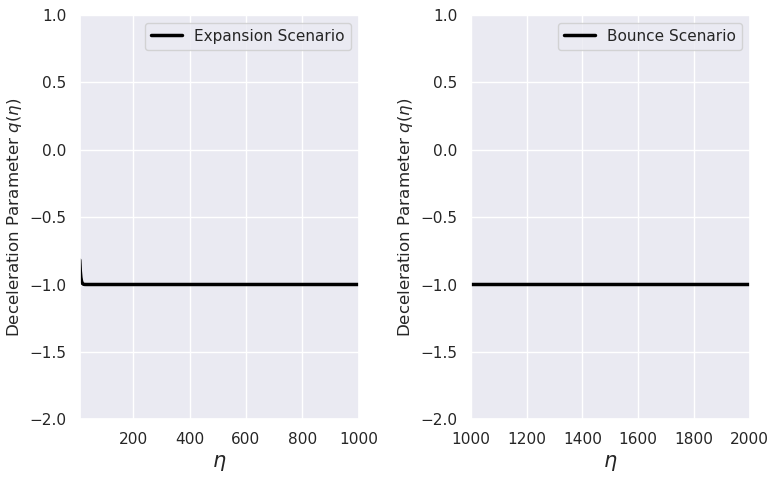}
    \caption{Deceleration parameter for the scenarios depicting the expansion and bounce solutions.}
    \label{deceleration}
\end{figure}

Figure~\ref{hubble_parameter} illustrates two representative cases drawn from Table~\ref{tab:inflationary_solutions}, corresponding to the expanding and bouncing solution regions, respectively. In both cases, the Hubble parameter remains constant throughout the interval during which the exponential growth occurs. Consistent with the $\Lambda$CDM expectation, Fig.~\ref{deceleration} confirms that $q \to -1$ for both families of solutions.

%-------------------------------------------------------------------------------------------------------

%-------------------------------------------------------------------------------------------------------

\section{The Quantum Model}
\label{sec:quantum_model}

We quantize the theory following Dirac's formalism \cite{Witt, Dirac}. This is done by promoting the variables $(x, T)$ to operators, together with their respective canonical momenta
\begin{equation}
    \hat{p_x} \longrightarrow -i \frac{\partial}{\partial x},\ \hat{p_T} \longrightarrow -i \frac{\partial}{\partial 
    T},
\end{equation}
\noindent where we use $\hbar = 1$. We now proceed to the canonical quantization of the model. The classical Hamiltonian constraint $(H = 0)$, where $(H)$ is the total Hamiltonian given in Eq.~(\ref{eq:H_dust}), is promoted to an operator equation acting on the wave function of the Universe $(\Psi(x,T))$. This yields the quantum constraint
\begin{equation}
    \hat{H} \Psi(x,T) = 0,
    \label{eq:wdw_general}
\end{equation}
where $\hat{H}$ denotes the Hamiltonian operator obtained via the standard replacement \(p_x \to -i\partial_x\) and \(p_T \to -i\partial_T\). Upon defining the new time variable \(\tau = -T\), the constraint equation above takes the form of the Wheeler--DeWitt (WdW) equation for the present model, namely
\begin{equation}
    \left[ -\frac{\partial^2}{\partial x^2} + m^2 \omega_k^2 \left( \frac{3x}{2} \right)^{2/3} - 2m\Lambda \left( \frac{3x}{2} \right)^{2} \right] \Psi(x,\tau) = 2im \frac{\partial \Psi(x,\tau)}{\partial \tau}
\label{eq:wdw_dust}
\end{equation}

\noindent We have computed the solution $\Psi(x,\tau)$ of the WdW equation numerically using finite diferences in the Crank-Nicolson scheme. Here we consider wave functions that satisfy the boundary conditions established by Hartle-Hawking \cite{HartleHawking1983},

\begin{equation}
\Psi(0, \tau) = \Psi(\infty, \tau) = 0.
\label{eq:HH_boundary}
\end{equation}
In the asymptotic limit \(x \to \infty\), the effective potential becomes negligible, and the wave function approaches a superposition of free propagating modes. The imposition of \(\Psi(x \to \infty, \tau) = 0\) selects the physically admissible (normalizable) solution, in which the transmitted wave packet disperses and its local probability density decays in both time and space. This corresponds to an open scattering problem characterized by a unidirectional probability flux.

If this asymptotic condition were not imposed, the problem would cease to be well-posed in the spectral sense of the Hamiltonian operator. From a numerical standpoint, this would be equivalent to solving the evolution equation on a finite domain with reflective boundary conditions. In such a scenario, the transmitted component would not be absorbed at infinity but would instead be reflected at the artificial boundary \(x_{\text{max}}\), returning to the barrier region. This reflection introduces additional terms into the probability flux, which can be interpreted as multiple successive incidences on the barrier, analogous to a geometric series of reflections.

\subsection{Integrated Tunneling Probability ($TP_{int}$)}

The first method for computing the tunneling probabilities in our quantum cosmological model requires calculating the likelihood of the Universe being found on the right-hand side of the potential barrier. To this end, we define the \textit{integrated tunneling probability} $TP_{\text{int}}$ \cite{alessandro3,gil1} as
\begin{equation}
TP_{\text{int}} = \frac{\int_{x_{\text{rtp}}}^{\infty} |\Psi(x, \tau_{\text{max}})|^2 \, dx}{\int_{0}^{\infty} |\Psi(x, \tau_{\text{max}})|^2 \, dx},
\label{eq:TP_int}
\end{equation}
where the upper limit of integration, in practice, is replaced by a suitably large numerical value, acting as a numerical infinity.

\begin{figure}
    \centering
    \includegraphics[scale=0.6]{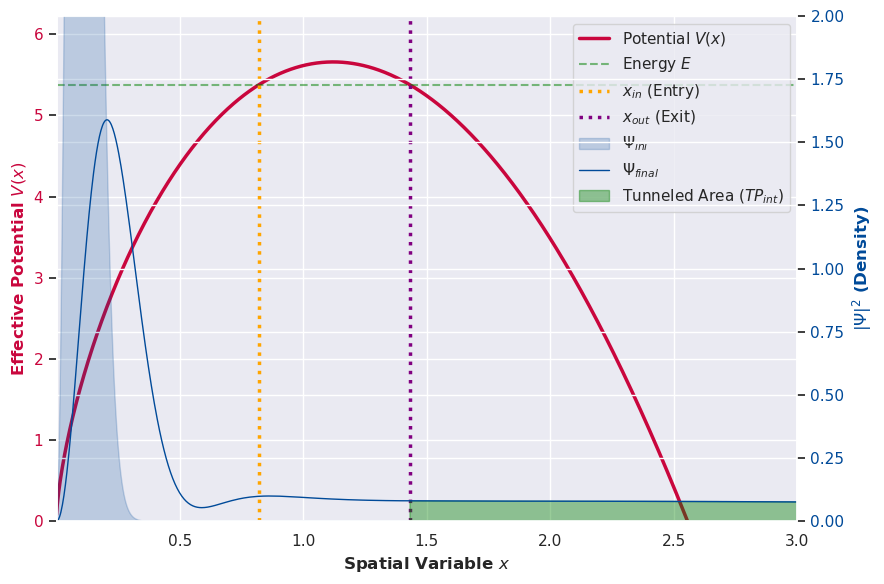}
    \caption{Effective potential $V_{\text{eff}}(x)$, initial $|\psi_0(x)|^2$, and evolved $|\Psi(x,\tau)|^2$ for $E = 5.37$. Vertical dashed lines mark the classical turning points $x_{\text{in}}$ and $x_{\text{out}}$; horizontal dashed line indicates the energy level. Snapshots at $\tau_{\text{in}}$ (barrier entry) and $\tau_{\text{out}}$ (barrier exit) are shown. The shaded region beyond $x_{\text{out}}$ represents the transmitted component contributing to $TP_{\text{int}}$ [Eq.~(\ref{eq:TP_int})].}
    \label{psi_sqr}
\end{figure}

\begin{figure}
    \centering
    \includegraphics[scale=0.4]{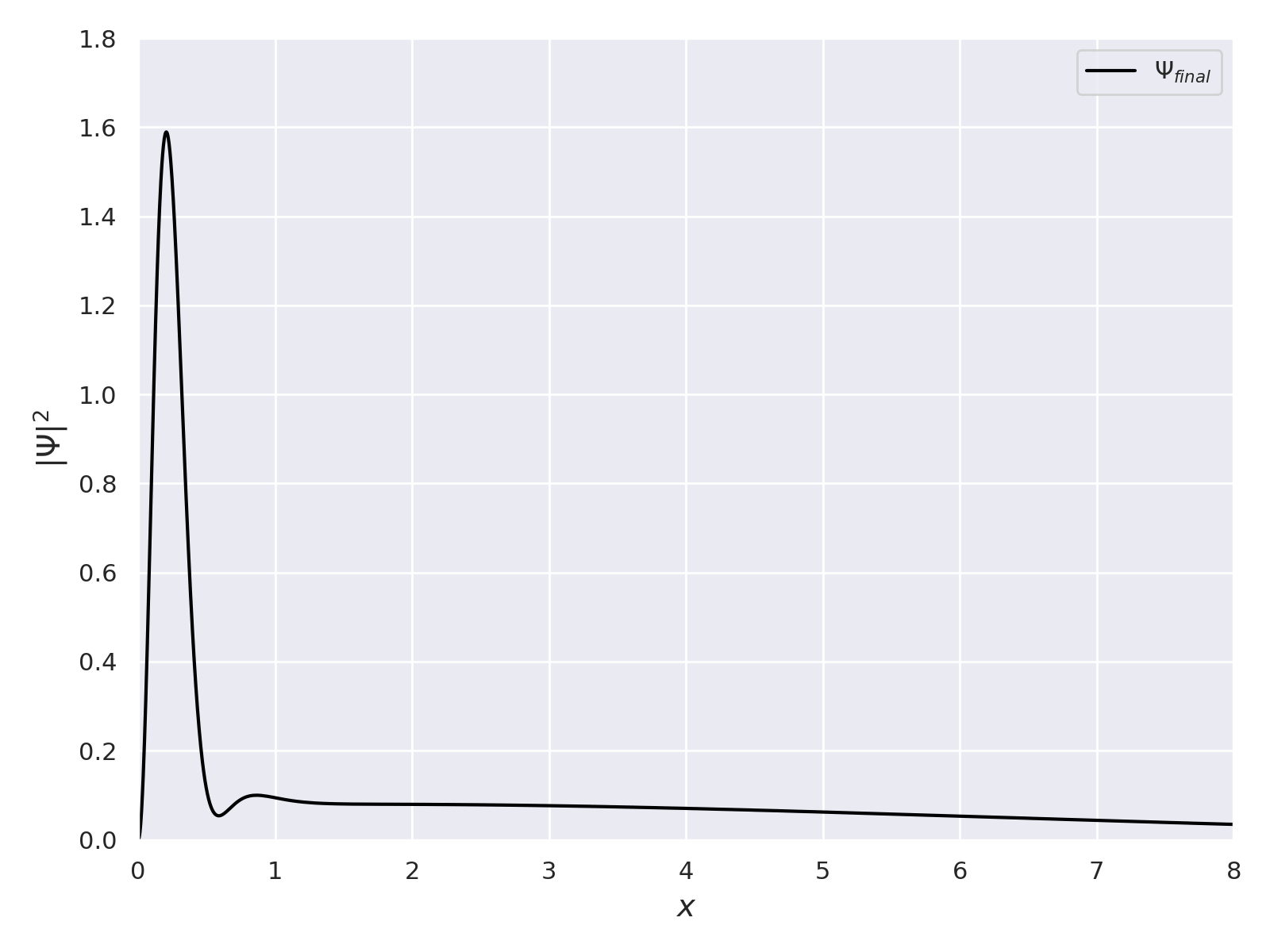}
    \caption{The square of the wave function $|\Psi|^2$ at $\tau_{max}$. A GIF showing the time evolution of $|\Psi|^2$ can be found \href{https://drive.google.com/file/d/1SYBsNKf3Dj0g0FBu2GkbK_MwxKHrLpHq/view?usp=sharing}{here}.}
    \label{psi_sqr_2}
\end{figure}

The initial state is taken to be a Gaussian-like wave packet centered on the left side of the barrier. Specifically, we adopt the following form for the wave function at $\tau = 0$ \cite{cond_ini_ref}:
\begin{equation}
\Psi(x,0) = \psi_0(x) = C \, x \exp\left(-\frac{2}{3} m E \, x^2\right),
\label{eq:initial_wavefunction}
\end{equation}
where the normalization constant is given by
\begin{equation}
C = 2.481612957 \left( \frac{m^3 E^3}{\pi} \right)^{1/4}.
\label{eq:norm_constant}
\end{equation}
Here, $E$ denotes the energy of the dust fluid. This choice of initial condition corresponds to a peaked distribution in the classically allowed region, which is then evolved forward in time according to the Wheeler--DeWitt equation in Eq.~(\ref{eq:wdw_dust}). Figure~\ref{psi_sqr} illustrates the initial probability density $|\psi_0(x)|^2$, the evolved probability density $|\Psi(x,\tau_{\text{max}})|^2$ where the value $\tau_{\text{max}}$ is defined as the time at which the wave packet reaches the numerical boundary ($|\Psi(x,\tau_{\text{max}})|^2$ is also shown by itself on Figure~\ref{psi_sqr_2}), and the effective potential $V(x)$ defined in Eq.~(\ref{eq:V_eff_dust}). The figure also indicates, by means of vertical dashed lines, the turning points $x_{\text{in}}$ and $x_{\text{out}}$, as well as the energy level, fixed at $E = 5.37$, which is represented by a horizontal line intersecting the potential.

\subsection{Expectation values in the Many Worlds Interpretation}

Proposed by Hugh Everett III in 1957, the \textit{many-worlds} (MW) interpretation of quantum mechanics \cite{Everett, Tipler} posits that the wave function of the Universe comprises a quantum superposition of all possible outcomes. In this framework, there is no collapse of the wave function, as occurs in the Copenhagen interpretation, where the act of measurement projects the state onto a single eigenstate through a non-unitary process. The MW interpretation naturally accommodates the context of quantum cosmological models described by the Wheeler--DeWitt equation in Eq.~(\ref{eq:wdw_dust}). Since this equation is independent of any external time parameter, it suggests that the classical notion of temporal evolution emerges only locally, within each branch of the wave function superposition. This framework provides a natural setting for addressing the problem of time and the interpretation of the quantum state of the Universe. In the MW interpretation the scale factor's behavior may be studied through the analysis of its expectation value given by 

\begin{equation}
\langle x(\tau) \rangle = \frac{\int_{0}^{\infty} \Psi^*(x, \tau) \, x \, \Psi(x, \tau) \, dx}{\int_{0}^{\infty} \Psi^*(x, \tau) \, \Psi(x, \tau) \, dx},
\label{eq:expectation_x}
\end{equation}

Using the initial condition given in Eq.~(\ref{eq:initial_wavefunction}), we numerically solve the Wheeler--DeWitt equation (\ref{eq:wdw_dust}) for the wave function $\Psi(x,\tau)$. From this solution, we then evaluate the expectation value $\langle x(\tau) \rangle$ defined in Eq.~(\ref{eq:expectation_x}). For the numerical integration, we adopt the following set of model parameters: $\Lambda = 1.0$, $\beta = 0.27$, and $\sigma = 0.9999$, for the model parameters mentioned in Eq.~(\ref{eq:theory_parameters}). The computational grid consists of $N_x = 15000$ spatial points and $N_\tau = 1300$ temporal points, with step sizes $\Delta x = 0.0033$ and $\Delta \tau = 0.001$ for the spatial and fluid variables, respectively. We used $x_{max} = 120$ and $\tau_{max} = 4.5$.

\begin{figure}
    \centering
    \includegraphics[scale=0.45]{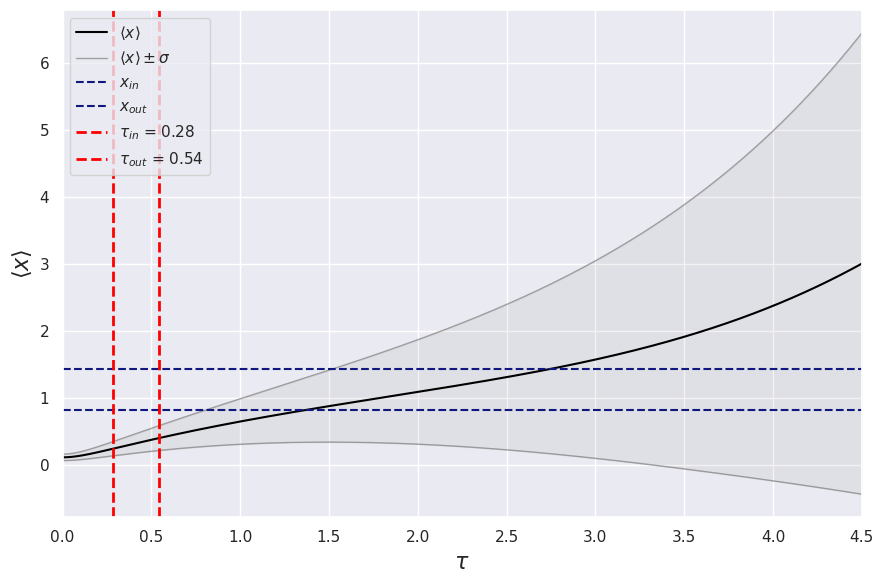}
    \caption{Time evolution of the expectation value $\langle x(\tau) \rangle$ as it traverses the potential barrier $V_{\text{eff}}(x)$ at energy $E = 5.37$. Horizontal dashed lines indicate the classical turning points $x_{\text{in}}$ and $x_{\text{out}}$; the instants $\tau_{\text{in}}$ and $\tau_{\text{out}}$, represented by vertical dashed lines, mark the wave packet's arrival at and departure from the barrier.}
    \label{exp_value}
\end{figure}

The resulting expectation value $\langle x(\tau) \rangle$ is represented by the central black line in Fig.~\ref{exp_value}. The adjacent grey lines correspond to $\langle x(\tau) \rangle \pm \varepsilon(\tau)$. These bands provide a measure of the quantum dispersion around the mean trajectory. $\varepsilon(\tau)$ denotes the standard deviation of $x$ computed from the probability density $|\Psi(x,\tau)|^2$: 

\begin{equation}
    \varepsilon(\tau) = \sqrt{\langle x(\tau)^2 \rangle - \langle x(\tau) \rangle^2}.
\end{equation}

\noindent The results displayed in Figure \ref{exp_value} demonstrate that neither the expectation value of the scale factor nor its associated uncertainties ever vanish. This reinforces the conclusion that this quantum model is free from the classical singularity.

\subsection{The de Broglie--Bohm Interpretation}

Originally proposed by Louis de Broglie and subsequently developed by David Bohm, the \textit{de Broglie--Bohm} (dBB) interpretation of quantum mechanics \cite{deBroglie1927, Bohm1952a, Bohm1952b} offers an alternative to the standard Copenhagen framework. In this formulation, particles possess well-defined positions at all times and are guided by a ``pilot wave'' represented by the wave function $\Psi(x,\tau)$. The evolution of the system is entirely deterministic: particles follow precise (though hidden) trajectories, with quantum randomness arising solely from ignorance of the initial conditions. The wave function does not undergo collapse but instead acts as a real physical field that actively influences the motion of particles. Non-locality is an intrinsic feature of the theory, reflecting the instantaneous correlations between entangled particles (as in the Einstein--Podolsky--Rosen thought experiment) and remaining fully consistent with Bell's theorem \cite{bell1964}.

This ontological commitment preserves a form of realism, as particles are assumed to exist independently of measurement, thereby circumventing paradoxes such as that of Schrödinger's cat. However, this realism comes at the cost of increased mathematical and conceptual complexity, largely due to the inherent non-locality of the theory.

Both the Many-Worlds and de Broglie--Bohm interpretations provide valid, although conceptually distinct, descriptions of quantum phenomena. With respect to the nature of the wave function, the MW interpretation denies any collapse; instead, all possible outcomes are equally real, each realized in a distinct branch of the universal wave function. The dBB interpretation similarly rejects wave-function collapse, but treats the wave function as an actual physical field that guides particle motion. Both frameworks are deterministic at the fundamental level, although probabilistic features may emerge in the MW picture due to the practical inaccessibility of the vast number of branching worlds.

Concerning the role of the observer, the MW interpretation posits that the observer branches into multiple versions, each inhabiting its own universe, whereas dBB maintains that the observer follows a single, well-defined trajectory guided by the pilot wave. The measurement problem is resolved in MW by abolishing collapse altogether and treating all possible outcomes as events that actually occur. In dBB, the problem is addressed through the introduction of hidden variables that determine the outcome of any given measurement. Regarding nonlocality, while the MW interpretation accepts the nonlocality implicit in standard quantum mechanics, dBB is explicitly nonlocal due to the global dependence of the particle dynamics on the entire wave function.

Each interpretation carries distinct philosophical implications. The MW interpretation entails the existence of a vast (and potentially infinite) multitude of equally valid parallel universes, each corresponding to a different outcome of every quantum event. By contrast, dBB proposes a fully realistic and deterministic framework, albeit one that inevitably relies on hidden variables.

Table \ref{tab:comparison_interpretations} makes a clear and useful comparison between Many-Worlds and DeBroglie-Bohm interpretations of Quantum Mechanics.

\begin{table}[htbp]
\centering
\small
\begin{tabular}{|p{5cm}|p{4.5cm}|p{4.5cm}|}
\hline
\textbf{Aspect of the theory} & \textbf{Many-Worlds (MW)} & \textbf{deBroglie--Bohm (dBB)} \\
\hline
\textbf{Nature of the Wave Function} & Wave function never collapses; every possibility is realized in a parallel universe & The wave function guides particles in a deterministic way \\
\hline
\textbf{Wave Function Collapse} & No collapse; all possibilities occur & No collapse; the pilot wave guides the particle \\
\hline
\textbf{Determinism} & Deterministic (but appears probabilistic due to the multiplicity of worlds) & Fully deterministic \\
\hline
\textbf{Realism} & Wave function is real, and the worlds exist objectively & Wave function and particles are both real \\
\hline
\textbf{Role of the Observer} & Observer does not cause collapse; it merely branches into different versions in distinct universes & Observer follows a specific trajectory guided by the pilot wave \\
\hline
\textbf{Measurement Problem} & Resolved by eliminating collapse and assuming that all possibilities are realized & Resolved by postulating hidden variables that determine the measurement outcome \\
\hline
\textbf{Nonlocality} & Accepts the nonlocality implicit in the Schr\"odinger equation & Explicitly nonlocal due to dependence on the global wave function \\
\hline
\textbf{Philosophical Implications} & Implies the existence of a vast (possibly infinite) number of parallel universes & Proposes a realistic and deterministic view of quantum mechanics, but requires hidden variables \\
\hline
\end{tabular}
\caption{Comparison between the de Broglie--Bohm and Many-Worlds interpretations of quantum mechanics.}
\label{tab:comparison_interpretations}
\end{table}

To employ the de Broglie--Bohm formulation of quantum mechanics, we begin by writing the wave function of the Universe in polar form,
\begin{equation}
    \Psi(x,\tau) = \mathcal{R}(x,\tau) \, e^{i \mathcal{S}(x,\tau)},
    \label{eq:polar_wave_function}
\end{equation}
where $\mathcal{R}(x,\tau)$ and $\mathcal{S}(x,\tau)$ are real-valued functions of $x$ and $\tau$. Here, $\mathcal{R}$ is the amplitude, given by $\mathcal{R} = |\Psi| = \sqrt{\Psi^* \Psi}$, and $\mathcal{S}$ is the phase, defined as $\mathcal{S} = \operatorname{Im}(\ln \Psi)$ or, equivalently, $\mathcal{S} = \arctan(\Psi_{\text{Im}} / \Psi_{\text{Re}})$. We work in natural units throughout, setting $c = \hbar = 1$. Substituting this ansatz into the Wheeler--DeWitt equation yields, in the standard Bohmian interpretation, the guidance relation for the canonical momentum,
\begin{equation}
    p_x = \frac{\partial \mathcal{S}}{\partial x},
    \label{eq:bohm_momentum}
\end{equation}
along with the quantum potential,
\begin{equation}
    Q(x,\tau) = -\frac{1}{\mathcal{R}(x,\tau)} \frac{\partial^2 \mathcal{R}}{\partial x^2}.
    \label{eq:quantum_potential}
\end{equation}

The Bohmian trajectory for the scale factor is then obtained from the momentum relation. Recalling that $p_x = m \, \dfrac{dx}{d\tau}$, we combine this with Eq.~(\ref{eq:bohm_momentum}) to obtain the guidance equation,
\begin{equation}
    \frac{dx}{d\tau} = \frac{1}{m} \frac{\partial \mathcal{S}}{\partial x}.
    \label{eq:traj_bohm}
\end{equation}
This first-order differential equation determines the evolution of the bohmian trajectory, with the quantum potential $Q(x,\tau)$ encoding the influence of the quantum regime on the classical trajectory.

\begin{figure}[!htpb]
    \centering
    \includegraphics[scale=0.45]{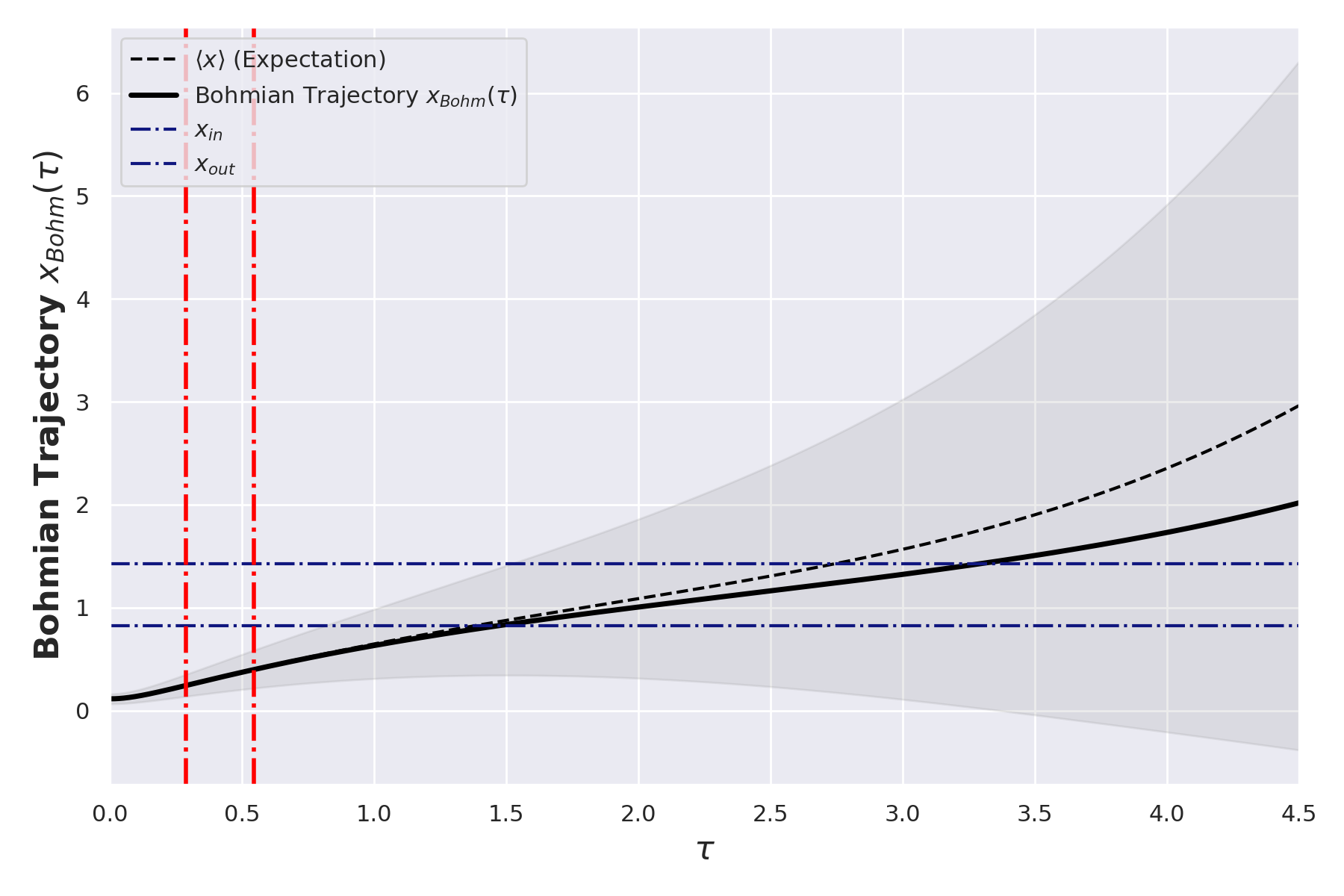}
    \caption{Bohmian trajectory $x_{Bohm}(\tau)$ (solid black) compared with the expectation value $\langle x \rangle (\tau)$ from the Many-Worlds interpretation (dashed black). The shaded region is delimited by $\langle x \rangle (\tau) \pm \varepsilon(\tau)$. Vertical dash-dotted lines indicate the moments when the wave function encounters and exits the barrier, while horizontal dash-dotted lines mark the classical turning points.}
    \label{bohm_traj_pic}
\end{figure}

Since an analytical expression for the wave function $\Psi(x,\tau)$ is not available, we adopt a fully numerical approach for this analysis, implemented in Python. The first step consists of extracting the amplitude $\mathcal{R}(x,\tau)$ and the phase $\mathcal{S}(x,\tau)$ from the time-evolved matrix representation of $\Psi(x,\tau)$. Subsequently, the derivative $\partial \mathcal{S} / \partial x$ is computed using the \texttt{gradient} function from the \textit{NumPy} library. To provide a smooth and differentiable right-hand side for the integration of the guidance equation, we construct a cubic spline interpolation of $\partial \mathcal{S}(x,\tau) / \partial x$ using the \texttt{RectBivariateSpline} routine. With this interpolated function, we numerically integrate Eq.~(\ref{eq:traj_bohm}) over $\tau$ employing the \texttt{solve\_ivp} function from the \textit{SciPy} library, selecting the explicit Runge--Kutta method of order 5(4) (RK45). The resulting solution yields the Bohmian trajectory $x_{\text{Bohm}}(\tau)$, which is displayed in Fig.~\ref{bohm_traj_pic}.

The initial condition for the Bohmian trajectory is chosen as the expectation value of the scale factor at $\tau = 0$, i.e., $x_B(0) = \langle x(\tau = 0) \rangle$, where $\langle x(\tau) \rangle$ is given by the solution to Eq.~(\ref{eq:expectation_x}) evaluated at the initial time. The Bohmian trajectory, obtained by numerically integrating Eq.~(\ref{eq:traj_bohm}) and displayed in Fig.~\ref{bohm_traj_pic}, corresponds to an expanding universe whose scale factor starts at a finite positive value and never reaches zero. This behaviour indicates that, analogously to the Many-Worlds interpretation, the present model is free from singularities at the quantum level. Notably, the Bohmian trajectory closely follows the expectation value, with only a slight deviation emerging after the wave packet exits the barrier at $\tau = \tau_{\text{out}}$.

\begin{figure}[htbp]
    \centering
    
    % Primeira linha
    \begin{subfigure}[b]{0.5\textwidth}
        \centering
        \includegraphics[width=\textwidth]{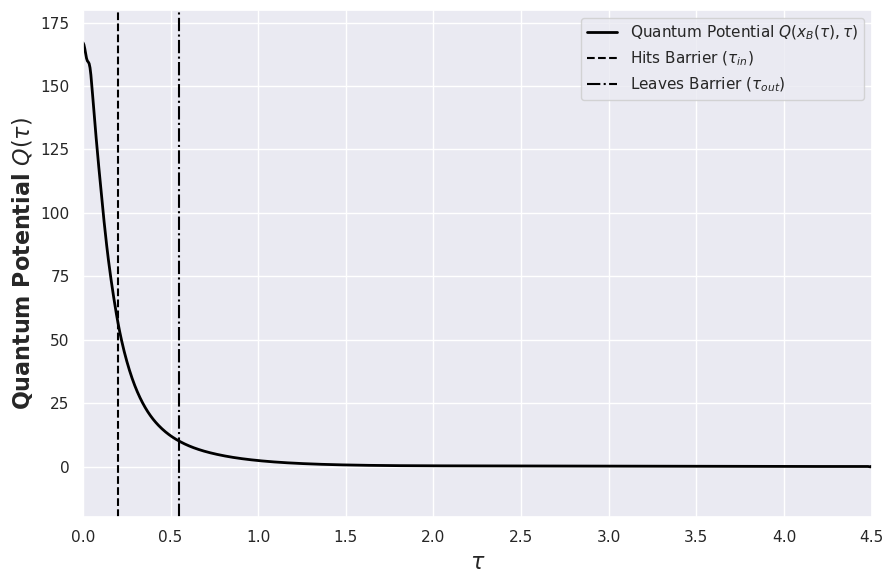}
        \caption{Quantum Potential $Q(\tau)$ in its full extension.}
        \label{quantum_pot_full}
    \end{subfigure}
    \hfill
    \begin{subfigure}[b]{0.45\textwidth}
        \centering
        \includegraphics[width=\textwidth]{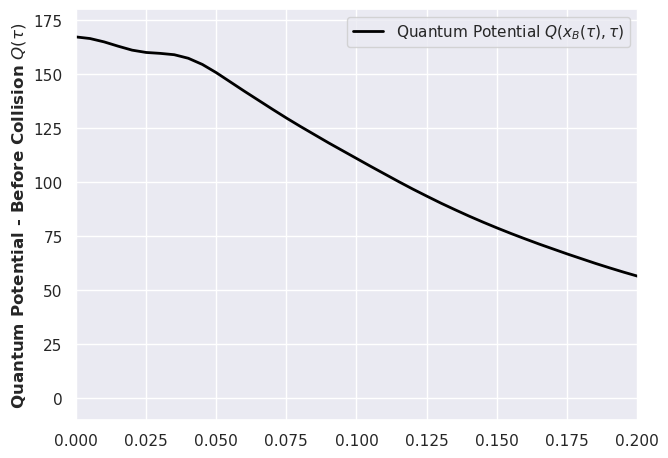}
        \caption{$Q(\tau)$ before hitting the barrier}
        \label{quantum_pot_1}
    \end{subfigure}
    
    \vspace{0.5cm} % Espaço vertical entre as duas linhas de gráficos
    
    % Segunda linha
    \begin{subfigure}[b]{0.45\textwidth}
        \centering
        \includegraphics[width=\textwidth]{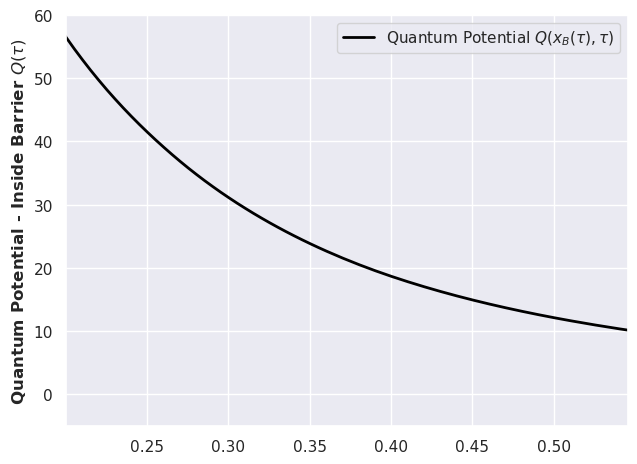}
        \caption{$Q(\tau)$ inside the barrier.}
        \label{quantum_pot_2}
    \end{subfigure}
    \hfill
    \begin{subfigure}[b]{0.45\textwidth}
        \centering
        \includegraphics[width=\textwidth]{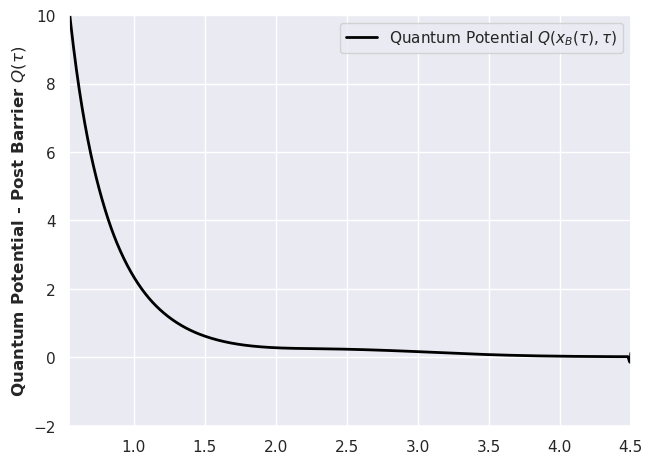}
        \caption{$Q(\tau)$ post barrier.}
        \label{quantum_pot_3}
    \end{subfigure}
    
    % Legenda e label gerais da figura inteira
    \caption{Quantum potential $Q\bigl(\tau\bigr)$ along the Bohmian trajectory. Three distinct phases are identified: (I) before the barrier, where the potential reaches its maximum shortly after the initial time; (II) inside the barrier, characterized by sinusoidal behaviour; and (III) after the barrier, where a beat-like modulation appears in the oscillations. The vertical lines indicate the barrier entry and exit times, $\tau_{\text{in}}$ and $\tau_{\text{out}}$.}
    \label{fig:quantum_pot_combined}
\end{figure}

We next compute the quantum potential $Q(x,\tau)$ by substituting the amplitude $\mathcal{R}(x,\tau)$, extracted from the numerical wave function, into Eq.~(\ref{eq:quantum_potential}). The resulting function is then evaluated along the previously obtained Bohmian trajectory $x_{\text{Bohm}}(\tau)$. The outcome of this procedure is shown in Fig.~\ref{quantum_pot_full}, where the quantum potential along the trajectory, $Q\bigl(x_{\text{Bohm}}(\tau), \tau\bigr) = Q\bigl(\tau\bigr)$, exhibits three qualitatively distinct regimes, corresponding to different phases of the dynamics:

\begin{enumerate}
    \item \textbf{Before the barrier} (Fig.~\ref{quantum_pot_1}): the quantum potential is extremely high and positive in this region, meaning the wave packet is tighly compressed in this region. The high curvature of the wave function implies a huge $Q(\tau)$ which acts as an intense repulsive force that prevents the classical singularity;
    \item \textbf{Inside the barrier} (Fig.~\ref{quantum_pot_2}): here the quantum potential exhibits a smooth decay, providing the necessary effective energy to sustain the tunneling process through the classically forbidden region;
    \item \textbf{After exiting the barrier} (Fig.~\ref{quantum_pot_3}): the quantum potential decays asymptotically to zero, reflecting the dispersion of the wave packet and the emergence of the classical cosmological regime.
\end{enumerate}

%-----------------------------------------------------------------------------------
%-----------------------------------------------------------------------------------

\section{The quantum birth of a classical Universe}
\label{sec:classical_quantum_transition}

In an attempt to test the limits of the present model, we investigate whether it can describe a Universe that originates as a purely quantum object and subsequently evolves into a classically describable regime. To this end, we examine both the quantum and classical phases of the dynamics, ensuring that the former is free from singularities while the latter is capable of accommodating an inflationary epoch.

Among a group of initial conditions which surely satisfies the boundary conditions, we have selected 
\begin{equation}
\psi_{\text{ini}}(x) = 2.481612957 \left( \frac{m^3 E^3}{\pi} \right)^{1/4} x \, \exp\left( -\frac{2}{3} m E \, x^2 \right),
\label{eq:psi_initial}
\end{equation}
which corresponds to a Gaussian-type wave packet peaked away from the origin, with the numerical coefficient chosen to ensure proper normalization. 

\begin{figure}[!htpb]
    \centering
    \includegraphics[scale=0.4]{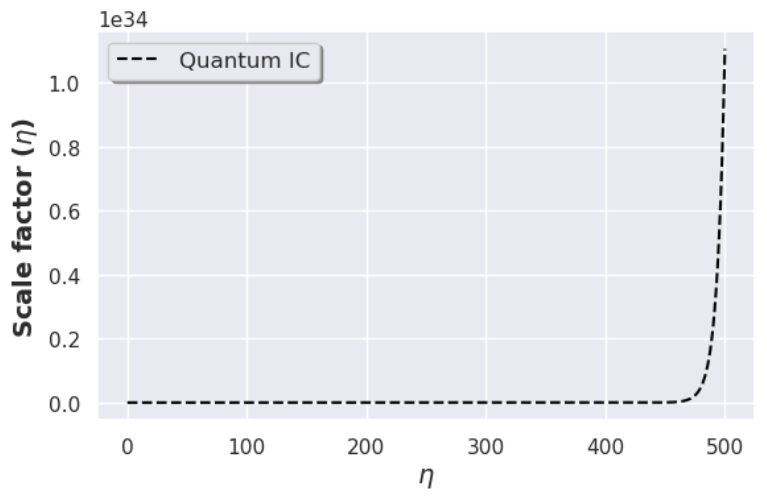}
    \caption{Evolution of the scale factor $x(\eta)$ obtained from the classical equation of motion, using initial conditions derived from the quantum exit point $\tau_{\text{out}}$. The solution exhibits a prolonged period of near-exponential growth, consistent with an inflationary regime.}
    \label{scale_factor_quantum_ic}
\end{figure}

\begin{figure}[!htpb]
    \centering
    \includegraphics[scale=0.4]{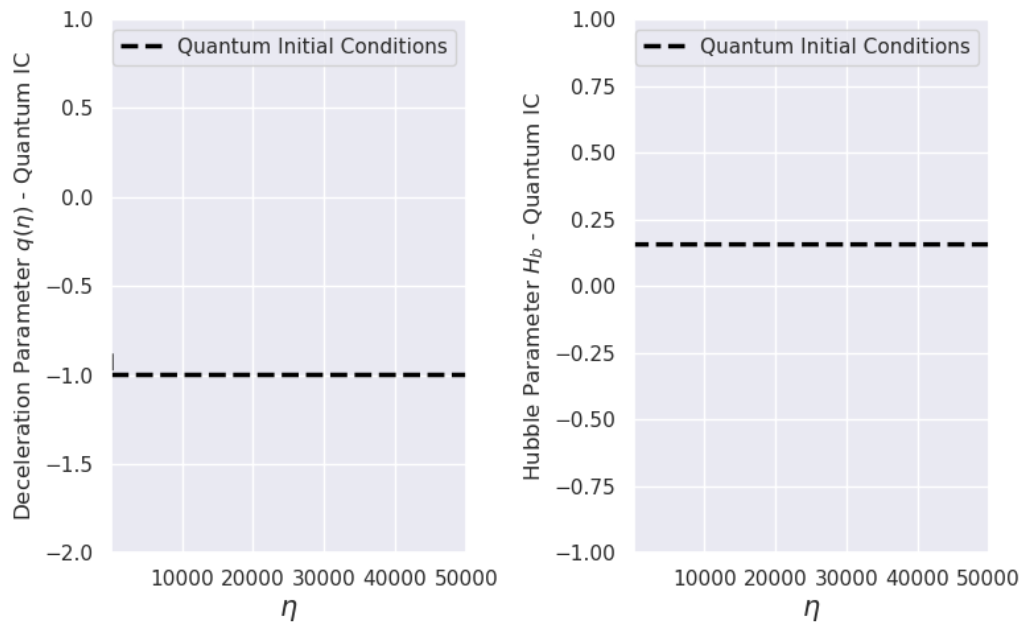}
    \caption{Hubble parameter $H_b(\eta)$ and deceleration parameter $q(\eta)$ as functions of conformal time $\eta$ for the quantum initial conditions case. During the inflationary epoch, $H_b$ remains constant at $\approx 0.25$, while $q = -1$.}
    \label{quantum_ic_inflation}
\end{figure}

We then substitute the initial condition of Eq.~(\ref{eq:psi_initial}) into Eq.~(\ref{eq:wdw_dust}) and evolve it numerically to obtain the full wave function $\Psi(x,\tau)$ of the Universe. With this solution at hand, we are then able to compute both the expectation value of the scale factor, $\langle x \rangle(\tau)$, via Eq.~(\ref{eq:expectation_x}), and the Bohmian trajectory $x_{\text{Bohm}}(\tau)$, via the guidance equation (\ref{eq:traj_bohm}). As discussed before, we determine the instants at which the wave function encounters the barrier, denoted $\tau_{\text{in}}$, and subsequently exits it, denoted $\tau_{\text{out}}$. Interpreting $\tau_{\text{out}}$ as the moment at which the Universe emerges from the barrier and transitions toward a classical regime, we construct a cubic spline interpolation of both the expectation value $\langle x \rangle(\tau)$ and its derivative $d\langle x \rangle/d\tau$ as functions of $\tau$. Evaluating these interpolated functions at $\tau_{\text{out}} = 0.5404157043879908$ yields
\begin{equation}
    \langle x \rangle(\tau_{\text{out}}) = 0.3999630542376294, \qquad 
    \left. \frac{d\langle x \rangle}{d\tau} \right|_{\tau = \tau_{\text{out}}} = 0.5913049622397701 .
    \label{eq:initial_conditions_x}
\end{equation}
These values are then adopted as the initial conditions for the classical equation of motion, Eq.~(\ref{eq:second_order_x}), thereby allowing us to evolve the scale factor forward in time from the exit of the barrier. As a first result, one can see in Fig.\ref{scale_factor_quantum_ic} that the scale factor goes through a considerable exponential growth. In Table~\ref{tab:inflationary_solutions}, we previously computed the scale factor ratio and the corresponding number of e-folds for a conformal-time interval satisfying $\eta_{\text{final}} / \eta_{\text{initial}} = 10^4$. Applying the same procedure to the present case, with $\eta_{\text{initial}} = \tau_{\text{out}} = 0.5404157043879908$ (the instant at which the wave function exits the barrier), we obtain a scale factor ratio of $x(\eta_{\text{final}}) / x(\eta_{\text{ini}}) = 1.2077 \times 10^{30}$, corresponding to a total of $69.2663$ e-folds. Finally, as shown in Fig.~\ref{quantum_ic_inflation}, the Hubble parameter remains constant throughout this epoch, while the deceleration parameter assumes the value $q = -1$.

%-----------------------------------------------------------------------------------
%-----------------------------------------------------------------------------------

\section{Dependence of tunneling probabilities on the parameters of the
theory}
\label{sec:tunneling_probs}

In this section, we compare the WKB tunneling probability, denoted $TP_{\text{WKB}}$, with the integrated tunneling probability $TP_{\text{int}}$, aiming to study their respective dependences on the energy $E$ and on the parameters $(\Lambda, \sigma, \beta)$.

\subsection{WKB Tunneling Probability $TP_{WKB}$}
Assuming that the solution to the Wheeler--DeWitt equation can be written as $\Psi(x,\tau) = \psi(x)e^{-iE\tau}$, the WdW equation takes the form

\begin{equation}
    \frac{d^2 \psi(x)}{dx^2} + \left[ 2m \left( E - V(x) \right) \right] \psi(x) = 0,
    \label{eq:wkb_schrodinger}
\end{equation}
in which $V(x)$ is the effective potential given in Eq.(~\ref{eq:V_eff_dust}). It is possible to use the WKB approximation to solve Eq.(\ref{eq:wkb_schrodinger}), and then use that solution to calculate the tunneling probabilities through the potential barrier $V(x)$ \cite{alessandro3,merzbacher,alessandro1,alessandro2}.  The solution to Eq.(\ref{eq:wkb_schrodinger}) has the form

\begin{equation}
\psi(x) = 
\begin{cases} 
\dfrac{A}{\sqrt{K(x)}} \exp\left(i \displaystyle\int_x^{x_l} K(x') \, dx'\right) + \dfrac{B}{\sqrt{K(x)}} \exp\left(-i \displaystyle\int_x^{x_l} K(x') \, dx'\right), & 0 \leq x \leq x_l, \\[1.2em]
\dfrac{C}{\sqrt{k(x)}} \exp\left(-\displaystyle\int_{x_l}^x k(x') \, dx'\right) + \dfrac{D}{\sqrt{k(x)}} \exp\left(\displaystyle\int_{x_l}^x k(x') \, dx'\right), & x_l \leq x \leq x_r, \\[1.2em]
\dfrac{F}{\sqrt{K(x)}} \exp\left(i \displaystyle\int_{x_r}^x K(x') \, dx'\right) + \dfrac{G}{\sqrt{K(x)}} \exp\left(-i \displaystyle\int_{x_r}^x K(x') \, dx'\right), & x_r \leq x < \infty,
\end{cases}
\label{eq:wkb_piecewise_updated}
\end{equation}

Here, $x_{\text{ltp}}$ and $x_{\text{rtp}}$ denote the left and right turning points of the barrier, respectively; these are defined as the points at which the condition $E = V(x)$ is satisfied, with $x_{\text{ltp}} \leq x_{\text{rtp}}$. The functions $K(x)$ and $k(x)$ are then defined as:
\begin{equation}
\begin{cases} 
K(x) = \sqrt{2m\bigl(E - V(x)\bigr)}, & \text{for } E > V(x), \\[0.8em]
k(x) = \sqrt{2m\bigl(V(x) - E\bigr)}, & \text{for } E < V(x).
\end{cases}
\label{eq:K_k_def}
\end{equation}

In the WKB solutions above, the coefficient $A$ is associated with the incident wave propagating from the origin toward the left of the barrier, while $B$ corresponds to the wave reflected by the barrier. Similarly, $F$ is identified with the transmitted wave emerging from the right side of the barrier, and $G$ denotes the wave incident from $x \to \infty$ propagating toward the right side of the barrier. These coefficients are related through the transfer matrix
\begin{equation}
\begin{pmatrix}
A \\
B
\end{pmatrix}
=
\begin{pmatrix}
2\theta + \frac{1}{2\theta} & i\left(2\theta - \frac{1}{2\theta}\right) \\[1.2em]
-i\left(2\theta - \frac{1}{2\theta}\right) & 2\theta + \frac{1}{2\theta}
\end{pmatrix}
\begin{pmatrix}
F \\
G
\end{pmatrix},
\label{eq:transfer_matrix}
\end{equation}
where the parameter $\theta$ is related to the height and length of the barrier,
\begin{equation}
\theta = \exp\left( \int_{x_{\text{ltp}}}^{x_{\text{rtp}}} k(x) \, dx \right).
\label{eq:theta_def}
\end{equation}

Assuming that no incident wave originates from the right-hand side of the barrier, we set $G = 0$ in the transfer matrix (Eq.~\ref{eq:transfer_matrix}). Under this condition, the transmission coefficient, which corresponds to the tunneling probability in the WKB approximation, is given by the ratio of the transmitted to the incident probability fluxes,
\begin{equation}
TP_{\text{WKB}} = \frac{|F|^2}{|A|^2}.
\label{eq:TP_WKB_final}
\end{equation}
This quantity represents the probability that a particle (or, in the cosmological context, the universe) tunnels through the classically forbidden region. From the transfer matrix in Eq.~(\ref{eq:transfer_matrix}), and imposing the condition $G = 0$, we obtain the relation
\begin{equation}
    A = \frac{1}{2} \left( 2\theta + \frac{1}{2\theta} \right) F.
    \label{eq:A_from_F}
\end{equation}
With this, one may obtain the relation
\begin{equation}
    TP_{\text{WKB}} = \frac{|F|^2}{|A|^2}
    = \frac{4}{\left( 2\theta + \frac{1}{2\theta} \right)^2}.
    \label{eq:TP_WKB_explicit}
\end{equation}
This expression makes explicit the dependence of the tunneling probability on the barrier parameters through the WKB exponent $\theta$, as defined in Eq.~(\ref{eq:theta_def}).

\subsection{Tunneling Probabilities as a function of the Energy $E$}
\begin{figure}[!htpb]
    \centering
    \includegraphics[scale=0.5]{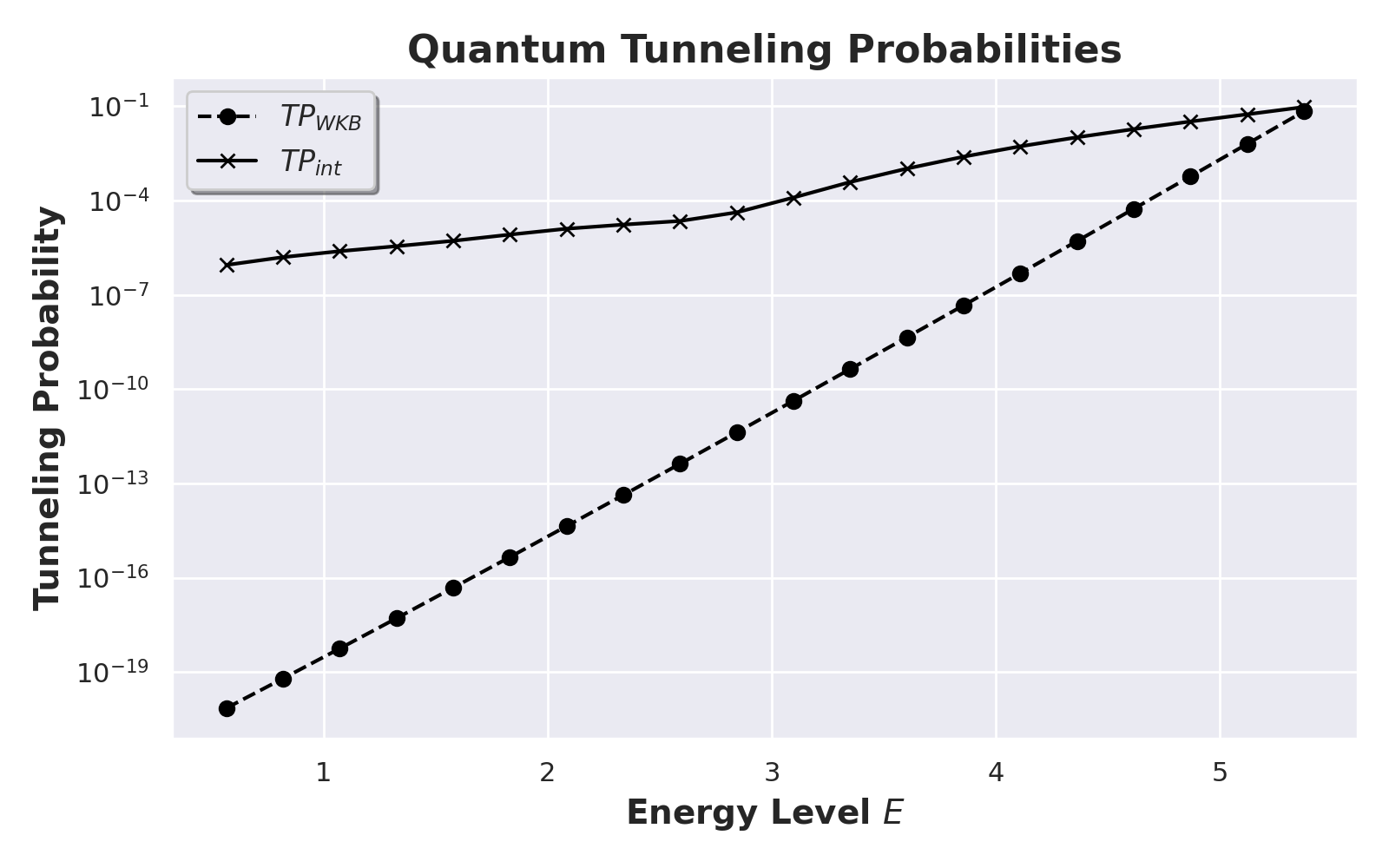}
    \caption{Comparison between $TP_{int}$ and $TP_{out}$ as functions of the fluid's energy. They tend to agree better for energies closer to the top of the barrier.}
    \label{tp_int_wkb_energy}
\end{figure}

Fixing the parameters $\sigma = 0.9999, \beta = 0.27, k=1, \Lambda = 1.0$ and $\alpha = 0$ for dust, the maximum value of the potential from Eq.(\ref{eq:V_eff_dust}) is approximately $V_{max} \approx 5.6579$. So, we varied the energy $E$ in an interval consisting of 20 equally spaced values, starting from $0.1V_{max}$ to $0.95V_{max}$. Fig~\ref{tp_int_wkb_energy}  compares the tunneling probabilities $TP_{int}$ and $TP_{WKB}$, which differ for lower values of energy but converge as the energy reaches the top of the barrier. 

Near the top of the barrier, the agreement between $TP_{\text{int}}$ and $TP_{\text{WKB}}$ can be understood by examining the effective dynamics of the wave packet in this regime. For energies $E$ close to the maximum of the potential, the barrier becomes comparatively narrow and shallow, which attenuates the typical exponential suppression associated with tunneling. In this situation, the assumptions underlying the validity of the WKB approximation are more readily satisfied, as the potential varies more gradually across the classically forbidden region. Moreover, the time required for the reflected component of the packet to return to the interaction region (after having traversed the barrier and been reflected by the infinite wall at $x = 0$) becomes sufficiently long. Consequently, during the time interval considered in the numerical integration, the dynamics are essentially governed by a single-incidence process on the barrier, which is precisely the scenario described by the WKB approximation.

In contrast, for energies $E$ substantially lower than the barrier peak, the system exhibits qualitatively distinct behaviour. In this regime, the effective barrier is broader and taller, rendering the WKB transmission probability extremely small as a consequence of the strong exponential damping. However, the presence of an infinite potential wall at $x = 0$ prevents the reflected component from dispersing, as would occur in a conventional scattering problem. Instead, the reflected wave returns and impinges upon the barrier anew, generating successive tunneling attempts. Each of these incidences contributes a small fraction of transmitted probability, and the cumulative effect of these multiple reflections can be interpreted as a sum of contributions analogous to a geometric series. Since the reflection coefficient is close to unity in this regime, this accumulation becomes significant, yielding a total tunneling probability substantially larger than the WKB prediction, which accounts only for a single incidence.

Hence, the discrepancy between the two results is directly tied to the effective open or confined character of the system. Near the top of the barrier, the problem effectively reduces to an open-domain scattering scenario, for which the WKB approximation is appropriate. For low energies, however, the system behaves as a quantum cavity, wherein successive reflections systematically enhance the transmission probability, thereby rendering the WKB estimate inadequate for an accurate description of the tunneling process.

\subsection{Tunneling Probabilities as a function of the Cosmological Constant $\Lambda$}

\begin{figure}[!htpb]
    \centering
    \includegraphics[scale=0.4]{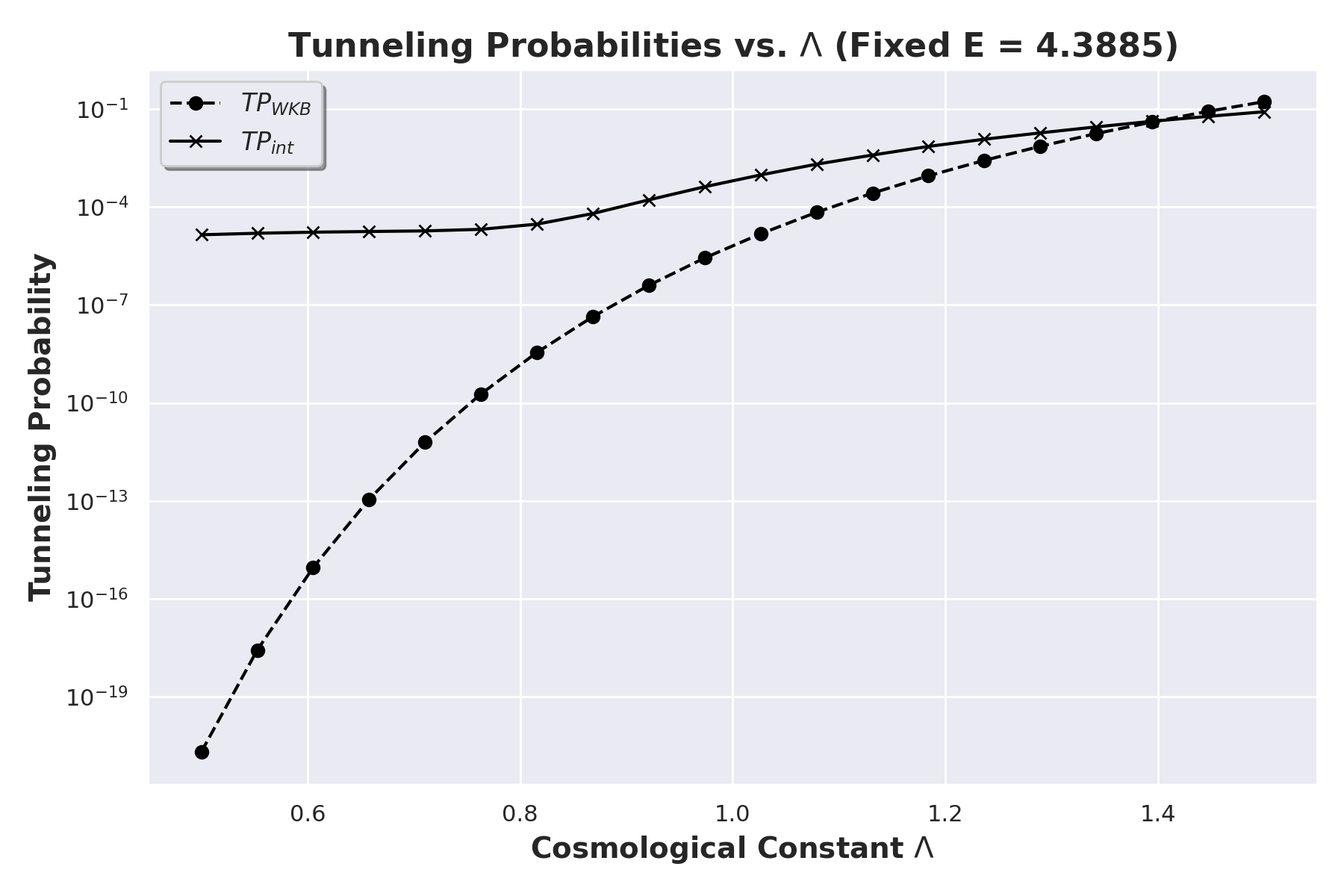}
    \caption{$TP_{int}$ and $TP_{out}$ VS $\Lambda$}
    \label{tp_int_wkb_lambda}
\end{figure}

By fixing $\sigma = 0.9999$, $\beta = 0.27$, $k = 1$, $\alpha = 0$, and $E = 4.3885$, and varying $\Lambda$ over the interval $[0.5, 1.5]$ in steps of $\Delta \Lambda = 0.05$, we obtain the behavior depicted in Fig.~\ref{tp_int_wkb_lambda}. The figure illustrates the dependence of the tunneling probabilities on the cosmological constant: as $\Lambda$ increases, both $TP_{\text{int}}$ and $TP_{\text{WKB}}$ increase and progressively converge toward the same value.

The maximum of the effective potential decreases with increasing $\Lambda$. This is the reason we have chosen $E$ to correspond to $95\%$ of the lowest barrier peak (i.e., the one associated with the largest value of $\Lambda$ considered). This criterion will be adopted consistently throughout the remainder of our analysis.

\subsection{Tunneling Probabilities as a function of the parameter $\sigma$}

\begin{figure}[!htpb]
    \centering
    \includegraphics[scale=0.4]{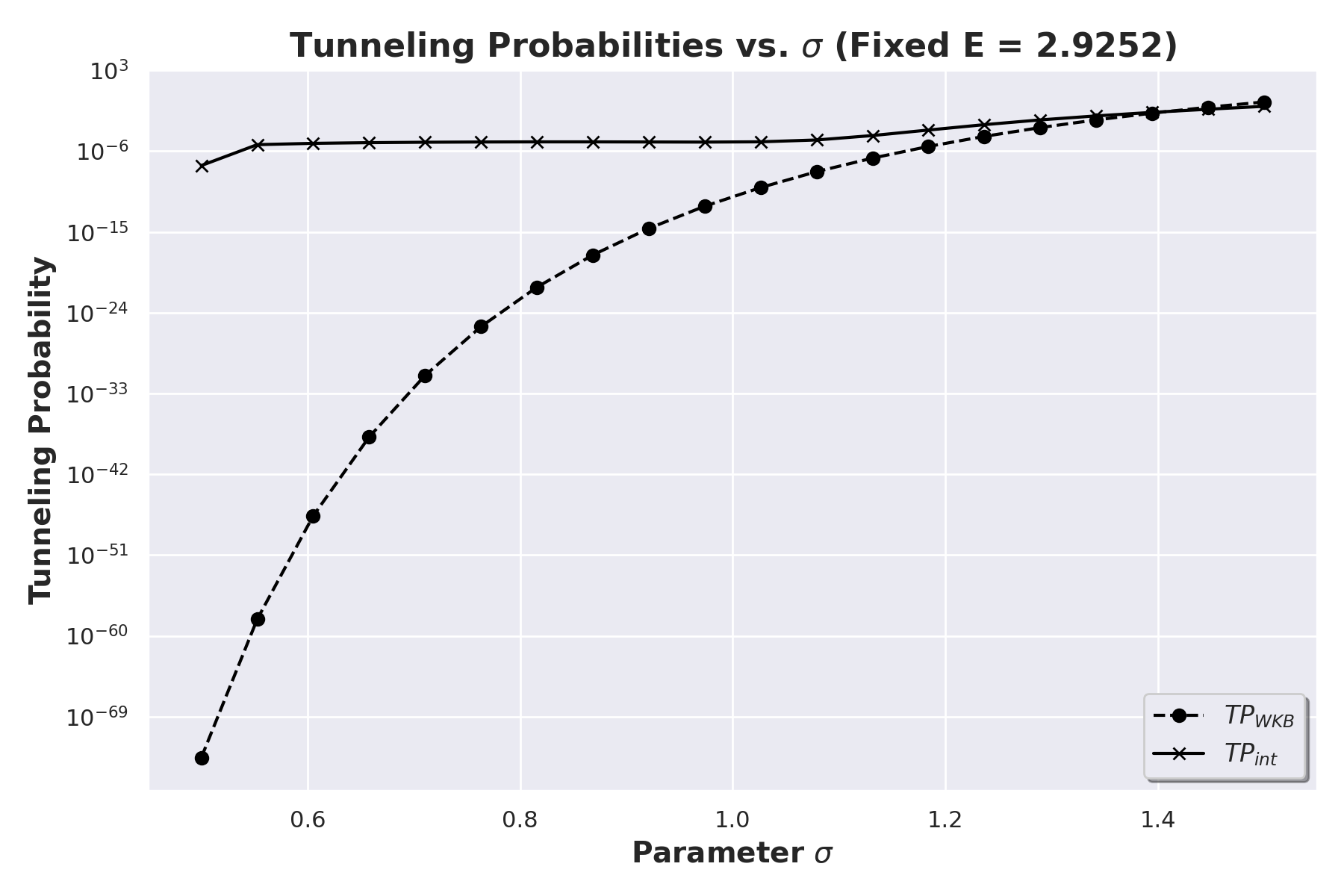}
    \caption{$TP_{int}$ and $TP_{out}$ VS $\sigma$}
    \label{tp_int_wkb_sigma}
\end{figure}

To investigate the behavior of the tunneling probabilities as a function of $\sigma$, we fixed $\beta = 0.27$, $k = 1$, $\Lambda = 1.0$, and $\alpha = 0$, with $E = 2.9252$, and varied $\sigma$ over the interval $[0.5, 1.5]$ in steps of $\Delta \sigma = 0.05$. We obtain the behavior depicted in Fig.~\ref{tp_int_wkb_sigma}. As $\sigma$ increases, both tunneling probabilities $TP_{\text{int}}$ and $TP_{\text{WKB}}$ increase and progressively converge toward the same value.

The maximum of the effective potential decreases with increasing $\sigma$. This is the reason we have chosen $E$ to correspond to $95\%$ of the lowest barrier peak.

\subsection{Tunneling Probabilities as a function of the parameter $\beta$}

\begin{figure}[!htpb]
    \centering
    \includegraphics[scale=0.4]{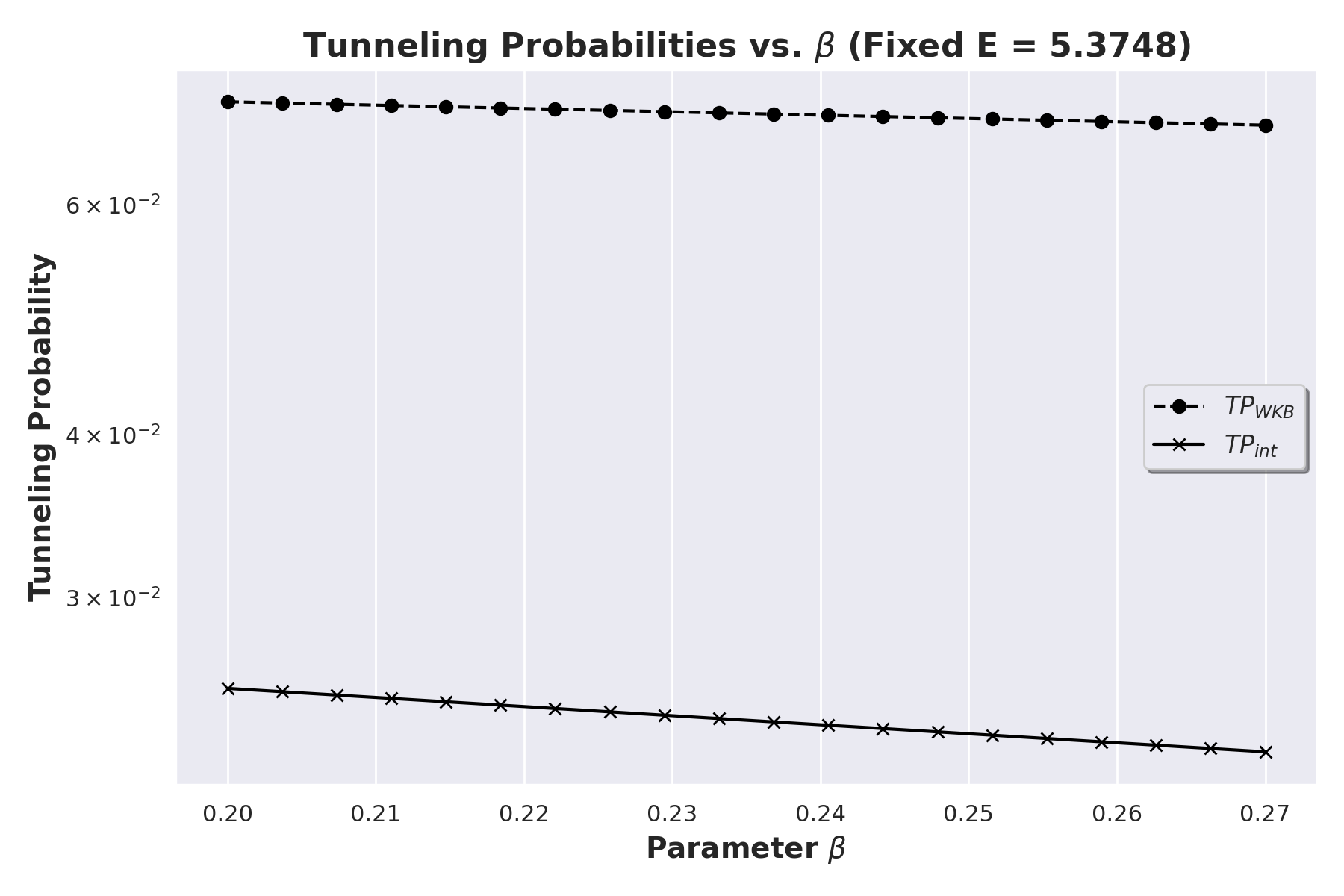}
    \caption{$TP_{int}$ and $TP_{out}$ VS $\beta$}
    \label{tp_int_wkb_beta}
\end{figure}

To investigate the behavior of the tunneling probabilities as a function of the parameter $\beta$, we fix $\sigma = 0.9999$, $k = 1$, $\Lambda = 0.05$, and $\alpha = 0$, and vary $\beta$ over the interval $[0.2, 0.27]$ in steps of $\Delta \beta = 0.0035$. The maximum of the effective potential changes with $\beta$; accordingly, we choose $E$ to correspond to $95\%$ of the lowest barrier peak, following the same criterion adopted in the previous analyses. As can be seen in Fig.~\ref{tp_int_wkb_beta}, the tunneling probabilities decrease monotonically with increasing $\beta$. This indicates that smaller values of $\beta$ yield higher tunneling probabilities.

\section{Conclusion}

In this paper, we explored the classical and quantum dynamics of an FLRW universe within Einstein-Æther gravity, considering a matter content of pressureless dust and a positive cosmological constant as dark energy. By constructing the minisuperspace Hamiltonian via Schutz’s variational formalism and analyzing the effective potential, we demonstrated that the classical dynamics naturally branches into four distinct classes of solutions: Big Crunch, expansion, bounce, and contraction. Notably, both the expansion and bounce branches capture the core features of standard inflation—delivering roughly sixty $e$-folds of accelerated expansion, a nearly constant Hubble parameter, and a deceleration parameter approaching $q=-1$.

\par Canonical quantization yielded the quantum version of the model, leading to the Wheeler-DeWitt equation, which we solved numerically via the Crank-Nicolson finite-difference scheme. The resulting quantum dynamics reveal that the Universe can emerge by tunneling through the effective potential barrier. We analyzed these results employing two complementary quantum mechanical frameworks. Within the Many-Worlds interpretation, the expectation value of the scale factor remains strictly positive throughout the evolution, thereby demonstrating the resolution of the classical singularity. The de Broglie-Bohm formulation yields an equivalent conclusion, as the corresponding Bohmian trajectories never reach a vanishing scale factor. Additionally, the Bohmian quantum potential clarifies the distinct dynamical regimes the Universe undergoes before, during, and after tunneling, the very process responsible for regularizing the singularity at the quantum level.

\par A pivotal outcome of this study is the establishment of a consistent quantum-to-classical transition. Instead of imposing classical initial conditions ad hoc, we extract them directly from the quantum dynamics the moment the wave packet exits the potential barrier. Seeded by these quantum-generated conditions, the subsequent classical evolution naturally unfolds into an inflationary phase of roughly 69 $e$-folds, characterized by a nearly constant Hubble parameter and a deceleration parameter approaching $q=-1$. Ultimately, this provides a continuous framework where a regular, singularity-free quantum universe transitions into a classical inflationary spacetime without requiring an inflaton scalar field.

\par Finally, we conducted a systematic analysis of the tunneling probability in terms of the dust energy, the Einstein-Æther couplings ($\beta$ and $\sigma$), and the cosmological constant ($\Lambda$). By comparing the integrated tunneling probability derived from the exact numerical Wheeler-DeWitt solution with the standard WKB approximation, we elucidated the physical origin of their agreements and discrepancies. Near the top of the potential barrier, where the scattering process essentially consists of a single wave-packet incidence, both approaches converge. At lower energies, however, multiple reflections within the quantum cavity significantly enhance the integrated tunneling probability compared to the standard WKB prediction.

\par Furthermore, increasing the cosmological constant ($\Lambda$) or the parameter ($\sigma$) enhances the tunneling probability, whereas increasing ($\beta$) suppresses it, revealing how Lorentz-violating couplings in Einstein-Æther gravity directly govern the quantum birth of the Universe. Ultimately, these results demonstrate that the present Einstein-Æther model provides a cohesive framework where quantum tunneling resolves the initial singularity, naturally generates classical initial conditions, and drives an inflationary phase consistent with standard cosmology. The absence of an inflaton field, coupled with the explicit quantum-to-classical bridge, suggests that Lorentz-violating gravity modifications offer a compelling alternative mechanism for the Universe's origin and early evolution. Future investigations will focus on the cosmological perturbations generated within this scenario, detailing their observational signatures against Cosmic Microwave Background (CMB) and Large-Scale Structure (LSS) data.

%%%%%%%%%%%%%%%%%%%%%%%%%%%%%%%%%%%%%%%%%%%%%%%%%%%%%%%%%%%%%%%%%%%%%%%%

	%%%%%%%%%%%%%%%%%%%%%%%%%%%%%%%%%%%%%%%%%%%%%%%%%%%%%%%%%%%%%%%%%%%%%%%%
	\section*{Author Contributions}
	%%%%%%%%%%%%%%%%%%%%%%%%%%%%%%%%%%%%%%%%%%%%%%%%%%%%%%%%%%%%%%%%%%%%%%%%
	
	All authors contributed to every stage of the paper, including the writing.
		
	%%%%%%%%%%%%%%%%%%%%%%%%%%%%%%%%%%%%%%%%%%%%%%%%%%%%%%%%%%%%%%%%%%%%%%%%
	\section*{Acknowledgments}
	%%%%%%%%%%%%%%%%%%%%%%%%%%%%%%%%%%%%%%%%%%%%%%%%%%%%%%%%%%%%%%%%%%%%%%%%
	
    G. A. Monerat thanks FAPERJ for partial financial support. G.A. Monerat and E. V. Corr\^{e}a Silva thanks Universidade do Estado do Rio de Janeiro, UERJ, for the Proci\^{e}ncia grant. A. Oliveira Castro Júnior thanks CAPES for financial support. This study was financed in part by the Coordena\c{c}\~{a}o de Aperfei\c{c}oamento de Pessoal de N\'{\i}vel Superior \"{} Brasil (CAPES)\"{} Finance Code 001. G. Oliveira-Neto thanks FAPEMIG (APQ-06640-24) for partial financial support. The Article Processing Charge (APC) for the publication of this research was funded by the Coordenação de Aperfeiçoamento de Pessoal de Nível Superior - Brasil (CAPES) (ROR identifier:00x0ma614). The authors thank Prof. Dr. Gustavo dos Santos Vicente for a critical reading of the manuscript and constructive suggestions. 
	%%%%%%%%%%%%%%%%%%%%%%%%%%%%%%%%%%%%%%%%%%%%%%%%%%%%%%%%%%%%%%%%%%%%%%%%
	\section*{Declarations}
	%%%%%%%%%%%%%%%%%%%%%%%%%%%%%%%%%%%%%%%%%%%%%%%%%%%%%%%%%%%%%%%%%%%%%%%%
	
	The authors declare that 
	\begin{enumerate}
		\item \textbf{Ethical Approval}: not applicable in our research shown here.
		\item \textbf{Competing interests}: not applicable in our research shown here.
		\item \textbf{Funding}: not applicable in our research shown here.
		\item \textbf{Availability of data and materials}: not applicable in our research shown here.
	\end{enumerate}
	
	%%%%%%%%%%%%%%%%%%%%%%%%%%%%%%%%%%%%%%%%%%%%%%%%%%%%%%%%%%%%%%%%%%%%%%%%


\begin{thebibliography}{99}
		%%%%%%%%%%%%%%%%%%%%%%%%%%%%%%%%%%%%%%%%%%%%%%%%%%%%%%%%%%%%%%%%%%%%%%%%

        % Referencias usadas em ordem
        
        \bibitem{penrose1965} Roger Penrose, Gravitational Collapse and Space-Time Singularities, \textit{Physical Review Letters} \textbf{14}, no. 3, 57--59 (1965).

        \bibitem{hawking1966} Stephen W. Hawking, The Occurrence of Singularities in Cosmology. I, \textit{Proceedings of the Royal Society A} \textbf{294}, 1439, 511--521 (1966).
        
        \bibitem{hawking1973} Stephen W. Hawking and George F. R. Ellis, \textit{The Large Scale Structure of Space-Time}, Cambridge University Press, Cambridge Monographs on Mathematical Physics series (1973).
        
        \bibitem{HawkingPenrose1970} Hawking, S. W. ; Penrose, R., "The Singularities of Gravitational Collapse and Cosmology", Proceedings of the Royal Society of London. Series A, Mathematical and Physical Sciences, Volume 314, Issue 1519, pp. 529-548
        
        \bibitem{DeWitt1967} B. S. DeWitt, Phys. Rev. 160, 1113 (1967).
        
        \bibitem{Kiefer2012} Claus Kiefer; Manuel Krämer, "Quantum Gravitational Contributions to the Cosmic Microwave Background Anisotropy Spectrum", Phys. Rev. Lett. 108, 021301 – Published 13 January, 2012. DOI: https://doi.org/10.1103/PhysRevLett.108.021301
        
        \bibitem{Ashtekar2011} Abhay Ashtekar; Parampreet Singh, "Loop quantum cosmology: a status report", Class. Quantum Grav. 28 213001, 2011. DOI 10.1088/0264-9381/28/21/213001
        
        \bibitem{Bojowald2008} Bojowald, M. "Loop Quantum Cosmology". Living Rev. Relativ. 11, 4 (2008). https://doi.org/10.12942/lrr-2008-4
        
        \bibitem{rovelli1990} C. Rovelli and L. Smolin, Loop Space Representation of Quantum General Relativity, \textit{Nuclear Physics B} \textbf{331}, 80--152 (1990). https://doi.org/10.1016/0550-3213(90)90019-A
        
        \bibitem{ashtekar2004} Abhay Ashtekar and Jerzy Lewandowski, Background independent quantum gravity: a status report, \textit{Classical and Quantum Gravity} \textbf{21}, No. 15 (2004).
        
        \bibitem{green1984} M. B. Green and J. H. Schwarz, Anomaly Cancellation in Supersymmetric D=10 Gauge Theory and Superstring Theory, \textit{Physics Letters B} \textbf{149}, 117--122 (1984).
        
        \bibitem{maldacena1998} J. Maldacena, The Large N Limit of Superconformal Field Theories and Supergravity, \textit{Advances in Theoretical and Mathematical Physics} \textbf{2}, No. 2, 231--252 (1998).
        
        \bibitem{ambjorn2000} J. Ambj{\o}rn, J. Jurkiewicz, and R. Loll, A Non-Perturbative Lorentzian Path Integral for Gravity, \textit{Physical Review Letters} \textbf{85}, 924--927 (2000).
        
        \bibitem{loll2020} R. Loll, Quantum Gravity from Causal Dynamical Triangulations: A Review, \textit{Classical and Quantum Gravity} \textbf{37}, 013002 (2020).
        
        \bibitem{reuter1998} M. Reuter, Nonperturbative Evolution Equation for Quantum Gravity, \textit{Physical Review D} \textbf{57}, 971--985 (1998).
        
        \bibitem{weinberg1980} Steven Weinberg, Ultraviolet Divergences in Quantum Theories of Gravitation, \textit{General Relativity: An Einstein Centenary Survey}, 790--831 (1980).
        
        \bibitem{dirac1958} P. A. M. Dirac, The Theory of Gravitation in Hamiltonian Form, \textit{Proc. R. Soc. Lond. A} \textbf{246}, 333--343 (1958).
        
        \bibitem{arnowitt2008} R. Arnowitt, S. Deser, and C. W. Misner, Republication of: The dynamics of general relativity, \textit{Gen. Relativ. Gravit.} \textbf{40}, 1997--2027 (2008).
        
        \bibitem{teitelboim1982} C. Teitelboim, Quantum Mechanics of the Gravitational Field, \textit{Physical Review D} \textbf{25}, 3159 (1982).
        
        \bibitem{isham1993} C. J. Isham, Canonical Quantum Gravity and the Problem of Time, In: L. A. Ibort and M. A. Rodr{\'i}guez (eds) \textit{Integrable Systems, Quantum Groups, and Quantum Field Theories}. NATO ASI Series, vol 409. Springer, Dordrecht (1993).
        
        \bibitem{Wheeler1968} Wheeler, J. A. "Superspace and the nature of quantum geometrodynamics Batelle Rencontres: 1967 Lectures in Mathematics and Physics ed C DeWitt and JA Wheeler." (1968).
        
        \bibitem{Halliwell1991} J.J.Halliwell; J.B.Hartle, "Wave functions constructed from an invariant sum over histories satisfy constraints", Phys. Rev. D 43, 1170 – Published 15 February, 1991. DOI: https://doi.org/10.1103/PhysRevD.43.1170
        
        \bibitem{Kiefer2024} Claus Kiefer et al 2024 J. Phys.: Conf. Ser. 2883 012008. DOI 10.1088/1742-6596/2883/1/012008
        
        \bibitem{Misner1969} Charles W. Misner, "Quantum Cosmology I", Phys. Rev. 186, 1319 – Published 25 October, 1969. DOI: https://doi.org/10.1103/PhysRev.186.1319
        
        \bibitem{Halliwell1988} Jonathan J. Halliwell, "Derivation of the Wheeler-DeWitt equation from a path integral for minisuperspace models", Phys. Rev. D 38, 2468 – Published 15 October, 1988. DOI: https://doi.org/10.1103/PhysRevD.38.2468
        
        \bibitem{Kuchar1992} Karel V. Kuchař, "Extrinsic curvature as a reference fluid in canonical gravity", Phys. Rev. D 45, 4443 – Published 15 June, 1992. DOI: https://doi.org/10.1103/PhysRevD.45.4443
        
        \bibitem{HartleHawking1983} Hartle, Hawking, "Wave function of the Universe", Phys. Rev. D 28, 2960 – Published 15 December, 1983.
        
        \bibitem{Vilenkin1982} Alexander Vilenkin, "Creation of universes from nothing", Physics Letters B, Volume 117, Issues 1–2. DOI: 10.1016/0370-2693(82)90866-8
        
        \bibitem{Vilenkin1984} Alexander Vilenkin, "Quantum creation of universes", Phys. Rev. D 30, 509(R) – Published 15 July, 1984. DOI: https://doi.org/10.1103/PhysRevD.30.509
        
        \bibitem{AlvarengaLemos1998} Flávio G. Alvarenga; Nivaldo A. Lemos, "Dynamical Vacuum in Quantum Cosmology", General Relativity and Gravitation, Volume 30, pages 681–694 (1998)
        
        \bibitem{PintoNeto2013} N Pinto-Neto; J C Fabris, "Quantum cosmology from the de Broglie–Bohm perspective",  Class. Quantum Grav. 30 143001, 2013. DOI 10.1088/0264-9381/30/14/143001
        
        \bibitem{Ashtekar2006} Abhay Ashtekar, "Space and Time: From Antiquity to Einstein and Beyond", Volume 11, pages 4–19 (2006)
        
        \bibitem{Linde1984} A D Linde, "The inflationary Universe", Rep. Prog. Phys. 47 925, 1984. DOI 10.1088/0034-4885/47/8/002 
        
        \bibitem{Coleman1980} Sidney Coleman; Frank De Luccia, "Gravitational effects on and of vacuum decay", Phys. Rev. D 21, 3305 – Published 15 June, 1980. DOI: https://doi.org/10.1103/PhysRevD.21.3305
        
        \bibitem{Feldbrugge2017} Job Feldbrugge1; Jean-Luc Lehners; Neil Turok, "No Smooth Beginning for Spacetime", Phys. Rev. Lett. 119, 171301 – Published 27 October, 2017. DOI: https://doi.org/10.1103/PhysRevLett.119.171301
        
        \bibitem{DiazDorronsoro2018} Dorronsoro, J. Diaz, et al. "Damped perturbations in the no-boundary state." Physical review letters 121.8 (2018): 081302.
        
        \bibitem{Monerat2007} Monerat, G. A., et al. "Dynamics of the early universe and the initial conditions for inflation in a model with radiation and a Chaplygin gas." Physical Review D—Particles, Fields, Gravitation, and Cosmology 76.2 (2007): 024017.
        
        \bibitem{Pedram2008} Pedram, P., et al. "Perfect fluid quantum Universe in the presence of negative cosmological constant." General Relativity and Gravitation 40.8 (2008): 1663-1681.
        
        \bibitem{Vakili2010} Vakili, Babak. "Classical and quantum dynamics of a perfect fluid scalar-metric cosmology." Physics Letters B 688.2-3 (2010): 129-136.
        
        \bibitem{OliveiraNeto2011} Neves, C., et al. "Canonical transformation for stiff matter models in quantum cosmology." International Journal of Modern Physics: Conference Series. Vol. 3. World Scientific Publishing Company, 2011.
        
        \bibitem{Alessandro2026} Castro Junior, A. Oliveira, et al. "Einstein–Aether primordial universe with radiation and dark energy." The European Physical Journal C 86.5 (2026): 565.
        
        \bibitem{Zeh1970} Zeh, H. Dieter. "On the interpretation of measurement in quantum theory." Foundations of Physics 1.1 (1970): 69-76.
        
        \bibitem{Joos1985} Joos, Eric, and H. Dieter Zeh. "The emergence of classical properties through interaction with the environment." Zeitschrift für Physik B Condensed Matter 59.2 (1985): 223-243.
        
        \bibitem{KieferPolarski1998} Kiefer, Claus, and David Polarski. "Emergence of classicality for primordial fluctuations: Concepts and analogies." Annalen der Physik 510.3 (1998): 137-158.
        
        \bibitem{Kiefer2009} Kiefer, Claus, and David Polarski. "Why do cosmological perturbations look classical to us?." Advanced science letters 2.2 (2009): 164-173.
        
        \bibitem{Halliwell1999} Halliwell, J. J. "Two approaches to coupling classical and quantum variables." International Journal of Theoretical Physics 38.11 (1999): 2969-2986.
        
        \bibitem{Jacobson2001} Jacobson, Ted, and David Mattingly. "Generally covariant model of a scalar field with high frequency dispersion and the cosmological horizon problem." Physical Review D 63.4 (2001): 041502.
        
        \bibitem{Jacobson2004} Jacobson, Ted, Stefano Liberati, and David Mattingly. "Astrophysical bounds on Planck suppressed Lorentz violation." Planck Scale Effects in Astrophysics and Cosmology (2005): 101-130.
        
        \bibitem{Jacobson2008} Jacobson, Ted. "Einstein-aether gravity: Theory and observational constraints." CPT and Lorentz Symmetry. 2008. 92-99.
        
        \bibitem{Mattingly2005} Mattingly, James. "Is quantum gravity necessary?." The universe of general relativity. Boston, MA: Birkhäuser Boston, 2005. 327-338.
        
        \bibitem{Liberati2013} Liberati, Stefano. "Tests of Lorentz invariance: a 2013 update." Classical and Quantum Gravity 30.13 (2013): 133001.
        
        \bibitem{Eling2004} Eling, Christopher, and Ted Jacobson. "Static post-Newtonian equivalence of general relativity and gravity with a dynamical preferred frame." Physical Review D 69.6 (2004): 064005.
        
        \bibitem{Lim2005} Lim, Eugene A. "Can we see Lorentz-violating vector fields in the CMB?." Physical Review D—Particles, Fields, Gravitation, and Cosmology 71.6 (2005): 063504.
        
        \bibitem{Carroll2004} Trodden, Mark, and Sean M. Carroll. "Introduction to cosmology." Particle Physics And Cosmology: The Quest for Physics Beyond the Standard Model (s)(TASI 2002). 2004. 703-793.
        
        \bibitem{Zlosnik2007} Zlosnik, Tom G., Pedro G. Ferreira, and Glenn D. Starkman. "Modifying gravity with the aether: An alternative to dark matter." Physical Review D—Particles, Fields, Gravitation, and Cosmology 75.4 (2007): 044017.
        
        \bibitem{Donnelly2010} Donnelly, William, and Ted Jacobson. "Coupling the inflaton to an expanding aether." Physical Review D—Particles, Fields, Gravitation, and Cosmology 82.6 (2010): 064032.
        
        \bibitem{Audren2013} Audren, Benjamin, et al. "Conservative constraints on early cosmology with MONTE PYTHON." Journal of Cosmology and Astroparticle Physics 2013.02 (2013): 001-001.
        
        \bibitem{Jacobson2014} Jacobson, Ted. "Undoing the twist: The Hořava limit of Einstein-aether theory." Physical Review D 89.8 (2014): 081501.
        
        \bibitem{oost2018} J. Oost, S. Mukohyama, and A. Wang, Constraints on Einstein-\AE ther theory after GW170817, \textit{Physical Review D} \textbf{97}, 124023 (2018).
        
        \bibitem{abbott2017} B. P. Abbott \textit{et al.}, Multi-messenger Observations of a Binary Neutron Star Merger, \textit{The Astrophysical Journal Letters} \textbf{848}, L12 (2017).
        
        \bibitem{castrojunior2026} A. O. Castro Junior, G. A. Monerat, F. G. Alvarenga, G. de Oliveira Neto, F. R. Manh{\~a}es, V. J. Monteiro, and J. C. G. Tedesco, Cosmic Strings and Negative $\Lambda$: Probing the Infant Universe in Einstein-\AE ther Theory, \textit{Brazilian Journal of Physics} \textbf{56}, 102-1--102-12 (2026).

        \bibitem{nelson2026} Vitenti, Sandro Dias Pinto, et al. "Two-fluid quantum bouncing cosmology: Theoretical model." Physical Review D 114.2 (2026): 023533.
        
        \bibitem{schutz1970} Bernard F. Schutz, Perfect Fluids in General Relativity: Velocity Potentials and a Variational Principle, \textit{Physical Review D} \textbf{2}, no. 12, 2762--2773 (1970).
        
        \bibitem{schutz1971} Bernard F. Schutz, Hamiltonian Theory of a Relativistic Perfect Fluid, \textit{Physical Review D} \textbf{4}, no. 12, 3559--3566 (1971).
        
        \bibitem{lapchinskii1977} V. G. Lapchinskii and V. A. Rubakov, Quantum Gravitation: Quantization of the Friedmann Model, \textit{Theoretical and Mathematical Physics} \textbf{33}, no. 3, 1076--1084 (1977).
        
        \bibitem{monerat2025} G. A. Monerat, H. J. Brumatto, G. Oliveira-Neto, F. G. Alvarenga, E. V. Corr{\^e}a Silva, and A. L. B. Ribeiro, Non-singular birth of the universe: High-performance numerical solutions of the Wheeler-DeWitt equation, \textit{Physics Letters B} \textbf{868}, 139623 (2025).
        
        \bibitem{lyth2002} David H. Lyth and David Wands, Generating the curvature perturbation without an inflaton, \textit{Physics Letters B} \textbf{524}, Issues 1--2, 5--14 (2002).
        
        \bibitem{lyth2003} David H. Lyth, Carlo Ungarelli, and David Wands, Primordial density perturbation in the curvaton scenario, \textit{Physical Review D} \textbf{67}, 023503 (2003).
        
        \bibitem{turner1983} Michael S. Turner, Coherent Scalar-Field Oscillations in an Expanding Universe, \textit{Physical Review D} \textbf{28}, no. 6, 1243--1247 (1983).
        
        \bibitem{moroi2000} Takeo Moroi and Lisa Randall, Wino Cold Dark Matter from Anomaly-Mediated SUSY Breaking, \textit{Nuclear Physics B} \textbf{570}, 455--472 (2000).
        
        \bibitem{allahverdi2010} Rouzbeh Allahverdi, Robert Brandenberger, Francis-Yan Cyr-Racine, and Anupam Mazumdar, Reheating in Inflationary Cosmology: Theory and Applications, \textbf{60}, 27--51 (2010).
        
        \bibitem{amin2015} Mustafa A. Amin, Mark P. Hertzberg, David I. Kaiser, and Johanna Karouby, Nonperturbative Dynamics of Reheating after Inflation: A Review, \textit{International Journal of Modern Physics D} \textbf{24}, 01, 1530003 (2015).
        
        \bibitem{carr2020} Bernard Carr and Florian K{\"u}hnel, Primordial Black Holes as Dark Matter: Recent Developments, \textit{Annual Review of Nuclear and Particle Science} \textbf{70}, 355--394 (2020).
        
        \bibitem{guth1981} Alan H. Guth, Inflationary Universe: A Possible Solution to the Horizon and Flatness Problems, \textit{Physical Review D} \textbf{23}, no. 2, 347--356 (1981).
        
        \bibitem{linde1982} Andrei D. Linde, A New Inflationary Universe Scenario: A Possible Solution of the Horizon, Flatness, Homogeneity, Isotropy and Primordial Monopole Problems, \textit{Physics Letters B} \textbf{108}, 6, 389--393 (1982).
        
        \bibitem{linde1983} Andrei D. Linde, Chaotic Inflation, \textit{Physics Letters B} \textbf{129}, no. 3--4, 177--181 (1983).
        
        \bibitem{mukhanov2005} Viatcheslav Mukhanov, \textit{Physical Foundations of Cosmology}, Cambridge University Press (2005).
        
        \bibitem{baumann2011} Daniel Baumann, TASI Lectures on Inflation, \textit{Physics of the Large and the Small, Proceedings of the Theoretical Advanced Study Institute in Elementary Particle Physics (TASI 2009)}, World Scientific, 523--686 (2011).

        \bibitem{Jacobson2008AEreport} Jacobson, Ted. "Einstein-aether gravity: A Status report." arXiv preprint arXiv:0801.1547 (2008).


        \bibitem{Schutz1} B. F. Schutz. Perfect fluids in General Relativity: Velocity potentials and a variational principle. Phys. Rev. D, 2:2762–2773, 1970.
		\bibitem{Schutz2} B. F. Schutz. Hamiltonian theory of a relativistic perfect fluid. Phys. Rev. D, 4:3559–3566, 1971.
        \bibitem{Furtado} F. G. Alvarenga, R. G. Furtado, R. Fracalossi, S. V. B. Gon\c{c}alves, Braz. J. Phys. 47, 96, 2016.

        \bibitem{campista2020} Campista et al., Can. J. of Phys. 98, 917 (2020)

        \bibitem{silva2009} Silva, EV Corrêa, et al. "Symplectic method in quantum cosmology." Physical Review D—Particles, Fields, Gravitation, and Cosmology 80.4 (2009): 047302.

         \bibitem{inflation1} Turner, Michael S. (2022-09-26). "The Road to Precision Cosmology". Annual Review of Nuclear and Particle Science. 72 (1): 1–35. DOI: https://doi.org/10.1146/annurev-nucl-111119-041046
        \bibitem{inflation2} Gibbons, Gary W., Stephen W. Hawking, and Stephen TC Siklos. "The Very early Universe: proceedings of the Nuffield workshop, Cambridge, 21 June to 9 July, 1982." Very Early Universe (1983).
        \bibitem{inflation3} Achúcarro, Ana, et al. "Inflation: theory and observations." arXiv preprint arXiv:2203.08128 (2022). DOI: https://doi.org/10.48550/arXiv.2203.08128

        \bibitem{WeinbergCosmology} S. Weinberg. {\it Cosmology}. New York: Oxford University Press, 2008.

        \bibitem{hubble1} Livio, Mario, and Adam G. Riess. "Measuring the Hubble constant." Physics Today 66.10 (2013): 41-47.
        \bibitem{hubble2} Hubble, Edwin. "A relation between distance and radial velocity among extra-galactic nebulae." Proceedings of the national academy of sciences 15.3 (1929): 168-173.
        \bibitem{deceleration} Camarena, David, and Valerio Marra. "Local determination of the Hubble constant and the deceleration parameter." Physical Review Research 2.1 (2020): 013028

        \bibitem{Witt} B. S. DeWitt, Phys. Rev. 160, 1113 (1967).
		\bibitem{Dirac} P. A. M. Dirac. Lectures on Quantum Mechanics. New York: Yeshiva University Press (1964).

        \bibitem{alessandro3} Castro Júnior, A. Oliveira, et al. "Birth of an Isotropic and Homogeneous Universe with a Running Cosmological Constant." Universe 11.9 (2025): 310. DOI: https://doi.org/10.3390/Universe11090310
        \bibitem{gil1} Acacio de Barros, J., et al. "Tunneling probability for the birth of an asymptotically de Sitter Universe." Physical Review D—Particles, Fields, Gravitation, and Cosmology 75.10 (2007): 104004. DOI: https://doi.org/10.1103/PhysRevD.75.104004

        \bibitem{cond_ini_ref} G.A. Monerat, F.G. Alvarenga, S.V.B. Gonçalves, G. Oliveira-Neto, C.G.M. Santos, E.V. Corrêa Silva, Eur. Phys. J. Plus, 137:117, https://doi.org/10. 1140/epjp/s13360-021-02316-9 (2022)

        \bibitem{Everett} H. Everett, III. The Many-Worlds Interpretation of Quantum Mechanics, ed. by B. S. DeWitt and N. Graham (Princeton University Press, Princeton, 1973).
		\bibitem{Tipler} F. J. Tipler. Phys. Rep. 137, 231, 1986.

        \bibitem{deBroglie1927} De Broglie, Louis. "La mécanique ondulatoire et la structure atomique de la matière et du rayonnement." Journal de Physique et le Radium 8.5 (1927): 225-241.
        
        \bibitem{Bohm1952a} Bohm, David. "A suggested interpretation of the quantum theory in terms of" hidden" variables. I." Physical review 85.2 (1952): 166.
        
        \bibitem{Bohm1952b} Bohm, David. "Reply to a criticism of a causal re-interpretation of the quantum theory." Physical Review 87.2 (1952): 389.

        \bibitem{bell1964} Bell, John S. "On the einstein podolsky rosen paradox." Physics Physique Fizika 1.3 (1964): 195.

        \bibitem{merzbacher} Merzbacher, E.: Quantum Mechanics. 3rd ed. (John Wiley and Sons, Inc., New York, 1998), Chap. 7
        \bibitem{alessandro1}Oliveira Castro Júnior, A., G. Oliveira-Neto, and G. A. Monerat. "Primordial dust Universe in the Hořava–Lifshitz theory." Modern Physics Letters A 39.23n24 (2024): 2450112. DOI: https://doi.org/10.1142/S0217732324501128
        \bibitem{alessandro2} Castro Júnior, A. Oliveira, G. Oliveira-Neto, and G. A. Monerat. "The initial moments of a Hořava-Lifshitz cosmological model." General Relativity and Gravitation 56.10 (2024): 125. DOI: https://doi.org/10.1007/s10714-024-03310-z
        
    %----------------------------------------------------------------------------

		
		
	\end{thebibliography}
\end{document}